\documentclass[aps,prb,twocolumn,groupaddress,floatfix,showpacs]{revtex4}
\usepackage{graphicx}
\usepackage{amssymb}
\usepackage{amsmath}
\usepackage{verbatim}

\ifx\pdftexversion\undefined
\usepackage[dvips]{hyperref}
\else
\usepackage{hyperref}
\fi
\hypersetup{
  colorlinks = true, linkcolor = blue
}

\begin{document}

\title{Two-particle entanglement in capacitively coupled Mach-Zehnder
interferometers}

\author{A.A.\ Vyshnevyy$^{\,a}$, A.V.\ Lebedev$^{\,b}$,
G.B.\ Lesovik$^{\,c}$, and G.\ Blatter$^{\,b}$}

\affiliation{$^{a}$Department of general and applied physics, MIPT, 
Institutskii per.\ 9, 141700 Dolgoprudny, Moscow region, Russia}

\affiliation{$^{b}$Theoretische Physik, Wolfgang-Pauli-Strasse 27,
ETH Zurich, CH-8093 Z\"urich, Switzerland}

\affiliation{$^{c}$L.D.\ Landau Institute for Theoretical Physics,
RAS, 119334 Moscow, Russia}

\date{\today}

\begin{abstract}

We propose and analyze a mesoscopic device producing on-demand entangled pairs
of electrons. The system consists of two capacitively coupled Mach-Zehnder
interferometers implemented in a quantum Hall structure. A pair of electron
wave-packets is injected into the chiral edge states of two (of the four)
incoming arms; scattering on the incoming interferometers splits the
wave-packets into four components of which two interact. The resulting
interaction phase associated with this component leads to the entanglement of
the state; the latter is scattered at the outgoing beam splitter and analyzed
in a Bell violation test measuring the presence of particles in the four
outgoing leads. We study the two-particle case and determine the conditions to
reach and observe full entanglement. We extend our two-particle analysis to
include the underlying Fermi seas in the quantum Hall device; the change in
shape of the wave-function, the generation of electron-hole pairs in the
interaction regime, and a time delay between the pulses all reduce the degree
of visible entanglement and the violation of the Bell inequality, effects
which we analyze quantitatively. We determine the device settings optimizing
the entanglement and the Bell test and find that violation is still possible
in the presence of the Fermi seas, with a maximal Bell parameter reaching
${\cal B} = 2.18 > 2$ in our setup.

\end{abstract}


\maketitle

\section{Introduction}

Quantum entanglement is a genuine property of quantum mechanics that has
attracted a lot of attention recently due to its potential usefulness as a
computational resource. As was first noticed by Bell~\cite{bell}, entangled
states can violate a certain type of inequality expressed in terms
of the correlation functions of measured outcomes. Later, Clauser and
coworkers~\cite{clauser} suggested a more transparent inequality which was
experimentally violated by photonic entangled states~\cite{aspect}.

The controlled creation and manipulation of {\it electronic}
entangled states in solid state devices is a challenging problem and
no experimental demonstration of a Bell inequality violation with
electrons is available so far. The main challenge lies in the fact
that, in contrast to photons, electrons are charged particles and
thus strongly interact with the electromagnetic environment, leading
to a fast decay of the coherence of the entangled state. On the
other hand, the Coulomb interaction allows one to easily create
entanglement among electrons.

During the last decade, several strategies have been suggested to create
entangled states of ballistically propagating electrons in mesoscopic devices.
The initial proposal was to use the generic spin-singlet entanglement of
Cooper pairs in a superconductor~\cite{lesovik_2001,recher_2001}, where the
constituent electrons of a Cooper pair are injected into two different normal
leads where they can propagate and thereby separate ballistically.  Recently,
the adiabatic splitting of Cooper pairs into normal conductors has been
demonstrated experimentally~\cite{hofstetter}. The constituents of
Cooper-pairs are entangled both in the orbital (energies) and spin degrees of
freedom.  The advantage of the spin-entanglement over the orbital entanglement
is that the spin coherence time can approach about $1$ ms in semiconductor
devices. However, the reliable demonstration of spin entanglement between
electrons in a Bell test~\cite{chtchelkatchev} requires the detection of spin
polarized currents at arbitrary polarization angles, that is still beyond the
level of present technology. Alternatively, one can use the energy degrees of
freedom of the injected Cooper pairs as an entangled variable~\cite{bayandin}.
The corresponding Bell test then requires the measurement of time-resolved
current correlators on the GHz scale in the frequency domain.

Other proposals for two-electron entanglement make use of interacting quantum
dot systems in the Coulomb blockade regime \cite{saraga_2003} or ballistic
electrons in integer quantum Hall devices~\cite{beenakker,samuelsson}. The
latter proposals involve the entanglement of the electrons' orbital degrees of
freedom and the corresponding Bell test requires the measurement of spinless
current correlators among different leads of the setup. However, in contrast
to the setups in Refs.\ \onlinecite{lesovik_2001,recher_2001,saraga_2003}, in
these schemes the entanglement is produced due to a postselection at the
moment of measurement and no interaction among electrons is required, see
Refs.\ \onlinecite{lebedev_04_05} for the discussion of the role of
projection.  In the postselection schemes, entangled states are produced with
less then $100\, \%$ efficiency, in contrast to the proposals using
interactions, where the entangled states are prepared deterministically by the
unitary evolution.

Another disadvantage in the above proposals is that they operate at finite
bias voltage where Cooper pairs or electrons are injected stochastically into
the device. As a result the Bell test requires the measurement of current
correlators at short times (of the order of the voltage time $\propto
\hbar/eV$). To go beyond the short time limit, one alternatively can operate
in the tunneling regime where electrons are injected rarely and one can
measure correlators at larger times $\propto \hbar/(T eV)$ ($T$ is the
transparency of the injecting lead). However, actual applications rely on the
controlled creation of entangled states on demand. It was suggested in Ref.\
\onlinecite{lebedev_pulse} to use Lorentzian voltage pulses to inject electron
pairs on demand; the Bell test then requires only the detection of the total
number of the transmitted particles through the different leads of the setup,
and no short-time analysis is needed. The controlled injection of individual
electrons into a device has been achieved recently, see Ref.\
\onlinecite{Feve}.

Considerable progress has been made in the fabrication of electronic
Mach-Zehner interferometers where electronic transport occurs through the edge
states of an integer quantum Hall state.  Experiments~\cite{heiblum,portier}
have uncovered an Aharonov-Bohm interference pattern at high visibility $\sim
62\, \%$. Later, the interference of two independent electrons coming from
different sources has been reported~\cite{neder}, reproducing the original
Hanbury Brown and Twiss experiment in optics with a visibility approaching
$70\, \%$.

Several theoretical proposals have been made to use coupled electronic
Mach-Zehnder interferometers to entangle two~\cite{kang,jordan} or three
independent electrons~\cite{vyshnevyy}. All these proposal deal with an
idealized situation where only two or three electrons are present in the
setup. However in real experiments~\cite{heiblum,portier}, a non-monotonic
dependence of the visibility factor has been observed at high bias voltages,
that cannot be explained within a framework of non-interacting electron
transport.  Thus, the electron interaction cannot be ignored and indeed its
proper accounting~\cite{sukhorukov} is required in order to explain the
experimental data.

The purpose of the present work is to analyse an entangling device for (two)
electrons consisting of two coupled Mach-Zehnder interferometers, where
electrons are injected on demand on top of the filled Fermi sea.  In contrast
to earlier proposals~\cite{kang,jordan,vyshnevyy}, here the injected electrons
are interacting not only with each other but also with the underlying Fermi
sea, leading to a parasitic entanglement with the surrounding electronic
environment. We formulate the corresponding Bell inequality test and find the
dependence of the Bell parameter ${\cal B}$ as a function of the Coulomb
interaction strength, the energy of the injected electrons, and the ratio
between the mutual and self-capacitances of the interacting leads.  As
expected, the interaction with the Fermi sea leads to a strong decoherence and
renders the Bell inequality violation more challenging (but still doable) as
compared with the idealized two-electron situation: injecting narrow
(high-energy) wave packets and for a mutual capacitance surpassing the
self-capacitance of the interacting leads we find that a Bell parameter ${\cal
B} \approx 2.18 > 2$ violating the Bell inequality ${\cal B} < 2$ can be
reached.

In the following, we first discuss our setup and define our goals, the
calculation of the maximally possible Bell parameter quantifying the
entanglement in the wave function and the implementation of the Bell test with
the expected outcomes for ${\cal B}$. In Section \ref{sec:2pBt} we discuss the
Bell test with only two particles in the setup. The analysis is extended to
include the underlying Fermi seas in the device in Section \ref{sec:BeFS} and
we conclude in Sec.\ \ref{sec:Con}. Technically, our analysis makes use of the
fact that the effects of capacitive interaction between chiral electrons can
be described by the action of voltage pulses \cite{lebedev11}.

\section{Bell setup}\label{sec:Bs}

\begin{figure}
\includegraphics[width=7.0cm]{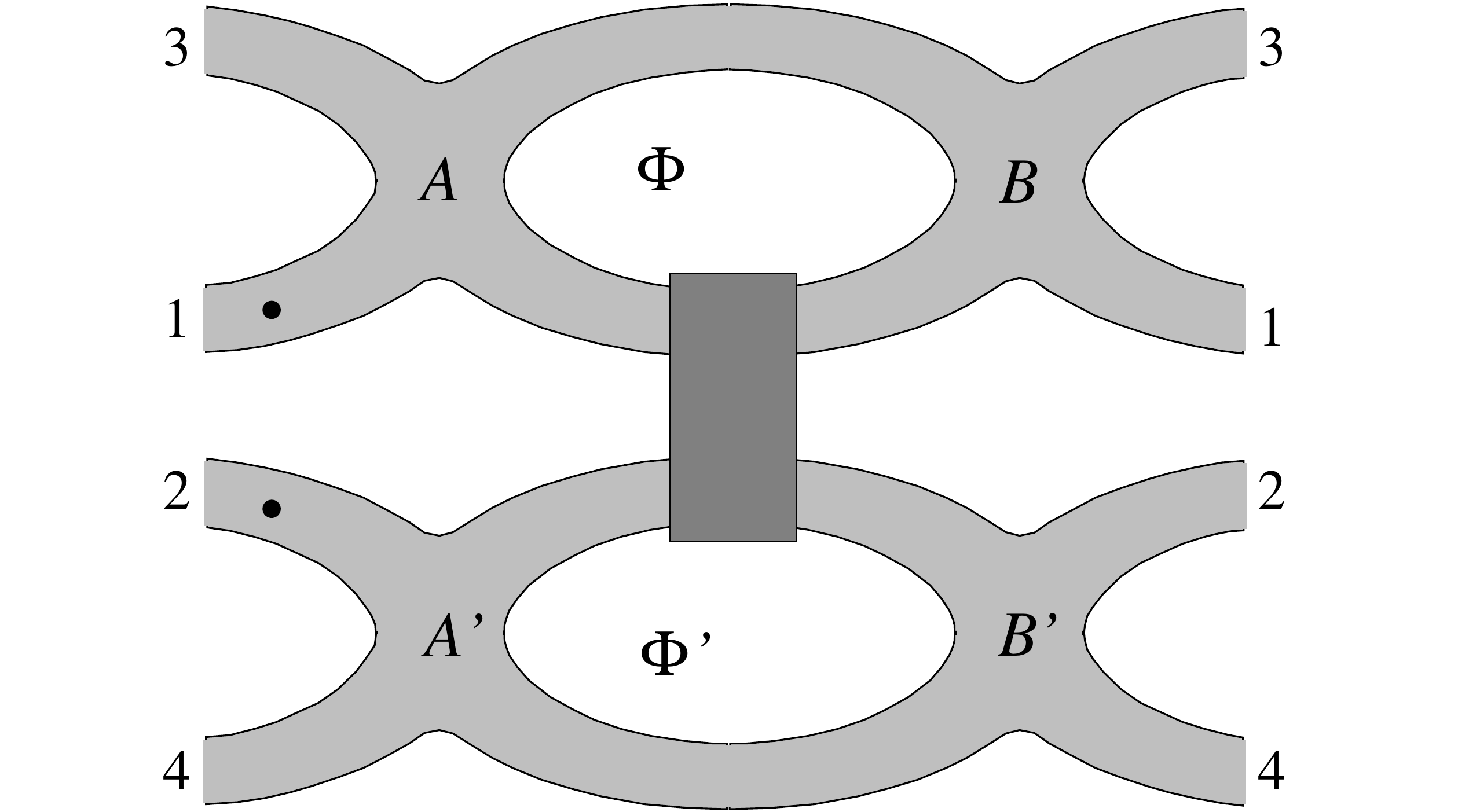}
\caption[]
{\label{fig:setup} Two Mach-Zehnder interferometers capacitively coupled via
the arms 1 and 2. Lorentzian-shaped pulses are injected into the left arms,
entangled through the Coulomb interaction, and analyzed in a Bell test on the
outgoing channels. In spite of the decoherence generated by the presence of
the Fermi sea, the Bell inequality can be violated, demonstrating the
possibility to generate `useful' entanglement correlations in this setup.}
\end{figure}

We consider two Mach-Zehnder interferometers, the upper with leads
$`1'$ and $`3'$ and a lower one with leads $`2'$ and $`4'$, where
the two adjacent arms $`1'$ and $`2'$ are capacitively coupled, see
Fig.~\ref{fig:setup}. Each interferometer consists of two
non-reflecting incoming and outgoing beam splitters $A$ ($A^\prime$)
and $B$ ($B^\prime$) for the upper (lower) interferometer. The beam
splitters $A$ and $B$ ($A^\prime$ and $B^\prime$) can be described by the
unitary transfer matrices $\hat{t}(\alpha), \hat{t}(\beta)$
($\hat{t}(\alpha^\prime), \hat{t}(\beta^\prime)$) in the lead basis
$\{`1',`3'\}$ ($\{`2',`4'\}$), with
\begin{equation}
      \hat{t}(\alpha)= \left[ \begin{array}{cc}
      \cos\alpha&-\sin\alpha\\ \sin\alpha&\cos\alpha
      \end{array} \right]
\end{equation}
parameterized by the angle $\alpha$ and corresponding expressions for the
other matrices.

We describe the interaction between the electrons propagating in the adjacent
leads $`1'$ and $`2'$ by the Hamiltonian
\begin{equation}
      \hat{H}_\mathrm{int} = \frac12\sum_{i,j\in\{1,2\}}
      E_{ij}\, \hat{\cal N}_i \hat{\cal N}_j,
      \label{eq:hint}
\end{equation}
where $\hat{\cal N}_i = \int dx\, \kappa_i(x) :\!\!\hat\Psi_i^\dagger(x)
\hat\Psi_i(x)\!\!:$ is the excess electron number in a finite region of 
lead $i$ defined by the coordinate kernel $\kappa_i(x)$ localized near the
origin. We also use the notation $:\!\hat{A}\!: = \hat{A} - \langle vac|
\hat{A}|vac\rangle$ with $|vac\rangle$ denoting the ground state of the
system. The coupling matrix $\hat{E}$ describes the coupling constants due to
the self-capacitances of the interferometer arms (diagonal terms $E_{11}$ and
$E_{22}$) and due to the mutual capacitance between the adjacent arms of the
different interferometers (off-diagonal terms $E_{12} = E_{21}$). For
simplicity, we assume no interaction in the other leads $`3'$ and $`4'$,
$E_{33} = E_{44} = 0$.

The excess electron charges are injected into the interferometers in a well
defined non-entangled quantum state.  Due to the particle interaction, the
outgoing state becomes orbitally entangled (in the particle number measured in
the outgoing leads) and this entanglement can be detected by performing a Bell
test and observing the degree of violation of the corresponding
CHSH-inequality. In the following, we formulate the Bell inequality in terms
of cross-correlators $\langle \hat{N}_i \hat{N}_j \rangle$ ($i\in\{`1',`3'\}$
and $j\in\{`2', `4'\}$) between the number of excess particles transmitted
through the different outgoing leads of the setup, $\hat{N}_i = \int dt\,
:\!\!\hat\Psi_i^\dagger(x,t) \hat\Psi_i(x,t)\!\!:$. These correlators can be
tuned by changing the magnetic fluxes $\Phi$ and $\Phi^\prime$ penetrating
each interferometer as well as the angles $\beta$ and $\beta^\prime$ of the
outgoing beam splitters (we will use symmetric splitters with $\alpha =
\alpha^\prime = \pi/4$ in the incoming leads). Defining the Bell correlator
\begin{equation}
      E_{\beta\beta^\prime}(\Phi,\Phi^\prime) =
      \frac{\langle (\hat{N}_1 - \hat{N}_3)(\hat{N}_2 - \hat{N}_4)\rangle}
      {\langle (\hat{N}_1 + \hat{N}_3)(\hat{N}_2 + \hat{N}_4)\rangle}
      \label{eq:bellcorr}
\end{equation}
and the corresponding Bell observable,
\begin{eqnarray}
      &&{\cal B} = E_{\beta\beta^\prime}(\Phi,\Phi^\prime) +
      E_{\beta\bar\beta^\prime}(\Phi,\bar\Phi^\prime)+
      \nonumber\\
      &&\qquad\qquad+
      E_{\bar\beta\beta^\prime}(\bar\Phi,\Phi^\prime) -
      E_{\bar\beta\bar\beta^\prime}(\bar\Phi,\bar\Phi^\prime),
      \label{eq:bellobservable}
\end{eqnarray}
one can formulate the two-party CHSH-inequality,
\begin{equation}
      |{\cal B}| \leq 2.
\end{equation}
The violation of the above inequality for some specific values of the magnetic
fluxes $\vec\Phi =\{\Phi,\Phi^\prime,\bar\Phi,\bar\Phi^\prime\}$ and angles
$\vec\beta=\{\beta,\beta^\prime,\bar\beta,\bar{\beta}^\prime\}$ quantifies the
entanglement shared between the two interferometers.  We will show that
violation (although not maximal in the full setup including the Fermi
surfaces) can be attained by changing only the magnetic fluxes at fixed
(optimal) values of the beam splitter angles, $\beta = \bar\beta$
and $\beta^\prime = \bar\beta^\prime$, reducing the number of adjustable
parameters, that might be helpful in a realistic experiment.

The actual value of the Bell parameter depends on the settings of the Bell
test, the magnetic fluxes $\vec\Phi$ and the adjustment of the outgoing beam
splitters defined by the angles $\vec\beta$: ${\cal B} = {\cal
B}(\vec\Phi,\vec\beta)$. In order to find the optimal setting of the Bell test
where the Bell parameter attains its maxima, one needs to maximize ${\cal
B}(\vec\Phi,\vec\beta)$ as a function of all tuning parameters in the test. It
has been shown in Ref.~\onlinecite{horodecky} that this optimization problem
can be done analytically by calculating the two-particle density matrix of the
entangled particles. We define the two-particle density matrix for our setup
by placing the outgoing beam splitters $B$ and $B^\prime$ into the asymptotic
region where $\kappa_{1,2}(x) \rightarrow 0$ with no interaction present (as
described by the limit $x\rightarrow +\infty$). In this region (before the
second beam splitters $B$ and $B^\prime$), the (many-particle) entangled
scattering state $|B B^\prime\rangle$ freely propagates in the positive
direction and we define the (reduced) density matrix
\begin{eqnarray} \label{eq:rhodef}
      [\hat\rho_{BB'}]_{ii^\prime jj^\prime} &\propto&  \int  dxdy\,
      \langle B B^\prime |
      \hat\Psi^\dagger_i(x) \hat\Psi^\dagger_{i^\prime}(y) \\
      \nonumber && \qquad \qquad \qquad \times
      \hat\Psi_{j^\prime}(y) \hat\Psi_j(x)|
      B B'\rangle,
\end{eqnarray}
where $\hat\Psi_i(x)$ denote the free single-particle electronic field
operators in the internal arms of the interferometers with $i,j \in
\{`1',`3'\}$ and $i^\prime,j^\prime \in \{`2',`4'\}$.  According to
Ref.~\onlinecite{horodecky} the maximal value of the Bell parameter is given
by,
\begin{equation}
      {\cal B}_\mathrm{max}(\hat\rho_{BB'}) = 2\sqrt{\lambda_1 +
      \lambda_2},
      \label{eq:horodecki}
\end{equation}
where $\lambda_1$ and $\lambda_2$ are the two largest eigenvalues of the
$3\times 3$ symmetric matrix $\hat{T}_{\rho_{BB'}}^\dagger
\hat{T}_{\rho_{BB'}}|$ with $[\hat{T}_\rho]_{nm} = \mbox{Tr}\,\{ \hat\rho\cdot
(\hat\sigma_n\otimes \hat\sigma_m)\}$ and where $\hat\sigma_n$,
$n,m\in\{\mathrm{x},\mathrm{y},\mathrm{z}\}$ are the Pauli matrices.

The phase accumulation due to magnetic fluxes $\Phi, \Phi^\prime$ and the
scattering on the outgoing beam splitters act as a unitary rotation of the
$4\times 4$ density matrix, 
\begin{eqnarray}\label{eq:density_out}
   \hat\rho_{BB'} \rightarrow \hat\rho_\mathrm{out} =
  \hat{U} (\Phi,\beta,\Phi^\prime,\beta^\prime) \, 
  \hat\rho_{BB'} \, \hat{U}^\dagger (\Phi,\beta,\Phi^\prime,\beta^\prime)
\end{eqnarray}
in the lead (or pseudo-spin) basis.  This unitary rotation does not change the
entanglement of the state since it involves only independent single-particle
rotations in each interferometer, $\hat{U}(\Phi,\beta,\Phi^\prime,
\beta^\prime) = \hat{U}(\Phi,\beta) \otimes \hat{U}(\Phi^\prime,\beta^\prime)$
where,
\begin{equation}
      \hat{U}(\Phi,\beta) = \hat{t}(\beta) \left( \begin{array}{cc}
      e^{i\Phi/2}&0\\0&e^{-i\Phi/2}
      \end{array} \right).
      \label{eq:Urot}
\end{equation}
Below, we will use Eq.\ (\ref{eq:horodecki}) in order to find the maximal
possible value of the Bell parameter in the device and Eq.\
(\ref{eq:density_out}) in order to calculate the Bell parameter
(\ref{eq:bellobservable}) from the correlators (\ref{eq:bellcorr}).

\section{Bell test with two electrons}\label{sec:2pBt}

We consider first the idealized situation where only two single-electron
wave-packets are injected into the incoming leads $`1'$ and $`2'$ and no other
electrons present in the system. The operator $\hat{f}_\alpha^\dagger = \int
dx\, f(x) \hat\Psi^\dagger_\alpha(x)$ creates a single-particle state with
wave-function $f(x)$ in lead $\alpha$. Then, the incoming state with two
wave-packets $f(x)$ and $g(x)$ in the leads `1' and `2' has the form,
$|\mathrm{in}\rangle = \hat{f}^\dagger_1 \hat{g}^\dagger_2\, |vac\rangle$.
After scattering on the incoming beam splitters $A$ and $A^\prime$, the
wave-packets are split between the different internal arms of the
interferometer. Before the wave-packets reach the interaction region the state
of the system is a non-entangled product state,
\begin{eqnarray}
      &&|AA^\prime\rangle = \bigl( \sin\alpha \, \hat{f}_3^\dagger +
      \cos\alpha \, \hat{f}_1^\dagger \bigr)
      \nonumber\\
      &&\qquad\qquad\times
      \bigl(\sin\alpha^\prime \, \hat{g}^\dagger_4 +
      \cos\alpha^\prime \, \hat{g}^\dagger_2 \bigr)
      |vac\rangle.
      \label{eq:psiA}
\end{eqnarray}

Depending on the path chosen by the electrons, the various
components of this product state evolve differently and the overall
two-particle state becomes non-separable with respect to the
different interferometers. For the empty vacuum state the
interaction effects appear only in the scattering component where
the electrons propagate through the adjacent arms $`1'$ and $`2'$.
Before the second beam splitters $B$ and $B^\prime$, the
two-particle state has the form,
\begin{eqnarray}
      &&|BB^\prime\rangle =
      \bigl[\cos\alpha\cos\alpha^\prime( \hat{S} \hat{f}^\dagger_1
      \hat{g}^\dagger_2)+
      \sin\alpha\sin\alpha^\prime
      (\hat{f}^\dagger_3 \hat{g}^\dagger_4)
      \nonumber\\
      &&\> + \sin\alpha\cos\alpha^\prime(\hat{f}_3^\dagger
      \hat{g}^\dagger_2)+
      \cos\alpha\sin\alpha^\prime(\hat{f}_1^\dagger
      \hat{g}^\dagger_4)\bigr] |vac\rangle,
      \label{eq:psiB2}
\end{eqnarray}
where $\hat{S}$ is the evolution operator corresponding to the interaction
Hamiltonian~(\ref{eq:hint}).  In order to find the scattering wave-function
$\hat{S} \hat{f}_1^\dagger \hat{g}_2^\dagger| vac\rangle$ of the two electrons
propagating in the interacting leads $`1'$ and $`2'$, we determine the
eigenvalues and eigenvectors $(\epsilon_1, \vec{e}_1)$ and $(\epsilon_2,
\vec{e}_2)$ of the coupling matrix $\hat{E}$ in the interaction
Hamiltonian~(\ref{eq:hint}) (note that the self-interaction terms have to be
set to zero in the two-particle problem discussed here, $E_{ii}=0$ for
$i=1,2$).  Introducing the electron number operators $\hat{n}_i = \sum_{j=1,2}
e_{ij} \hat{\cal N}_j$ ($e_{ij}$ is the $j$-th component of the vector
$\vec{e}_i$) one can rewrite the interaction Hamiltonian in the quadratic
form,
\begin{equation}\label{eq:quadr_form}
      \hat{H}_\mathrm{int} = \frac{\epsilon_1}2\, \hat{n}_1^2 +
      \frac{\epsilon_2}2\, \hat{n}_2^2.
\end{equation}
Next, we perform a Hubbard-Stratonovich transformation to decouple the charge
operators $\hat{n}_i$ (although not really necessary here, this procedure is
convenient when dealing with the many electron case later on, see Appendix).
The evolution operator then can be written in the form,
\begin{eqnarray}
      &&\hat{S} = \int Dz_1Dz_2\exp\Bigl[ \frac{i}2\int dt (\epsilon_1 z_1^2(t)
      + \epsilon_2z_2^2(t))\Bigr]
      \label{eq:HST}
      \\
      &&\qquad\times
      \hat{T}\exp\Bigl[-i\int dt\, ( \epsilon_1 z_1(t)\hat{n}_1(t) +
      \epsilon_2 z_2(t) \hat{n}_2(t)) \Bigr],
      \nonumber
\end{eqnarray}
where $z_{1,2}(t)$ are two real auxiliary fields obeying Gaussian statistics
with $\langle z_i(t) z_j(t^\prime)\rangle = (i/\epsilon_i)
\delta_{ij}\delta(t-t^\prime)$. This transformed evolution operator describes
the independent propagation of two electrons in different leads, each
subjected to a time-dependent scattering potential $V_i(x,t) =
V_i(t)\,\kappa_i(x)$, $i = 1,2$, with
\begin{equation}
      V_i(t) = \epsilon_1 z_1(t)\, e_{1i} + \epsilon_2
      z_2(t)\, e_{2i}.
\end{equation}
Having mapped the two-particle evolution to two independent single-particle
problems, we then solve the corresponding Schr\"odinger equations for the
chiral electron modes with linear spectrum $\epsilon = v\hbar k$. The
resulting two-particle wave-function $\Psi_{12}(x,y;t) = \langle x,y\,|\,
\hat{S} \hat{f}_1^\dagger \hat{g}_2^\dagger\,| vac\rangle$ emerging behind the
interaction region takes the form
\begin{eqnarray}
      &&\Psi_{12}(x,y;t) = f(x_t)g(y_t) \Bigl\langle 
      \exp\Bigl\{-i\!\int^t\!\! dt^\prime\,
      \label{eq:Psi12}\\
      &&\qquad
      \times 
      \bigl[V_1(t^\prime)\kappa_1(x_t+ vt^\prime) +
            V_2(t^\prime) \kappa_2(y_t +vt^\prime)
       \bigr]\Bigr\}\Bigr\rangle,
      \nonumber
\end{eqnarray}
where $x_t = x-vt$ and $y_t = y -vt$ are ballistically retarded
variables, and the average has to be taken with respect to the Gaussian
fields $z_i(t)$. Carrying out the averaging of the exponentials, we
find the result
\begin{equation}
      \Psi_{12}(x,y;t)= f(x_t)g(y_t) e^{-i\Phi_{12}(x,y)},
\end{equation}
\begin{equation}
      \Phi_{12}(x,y) = E_{12} \int\limits_0^\infty d\tau\,
      \kappa_1(x_\tau)\kappa_{2}(y_\tau).
\end{equation}
The phase $\Phi_{12}(x,y)$ describes both the energy exchange and the
deformation of the wave-functions $f(x)$ and $g(y)$ due to the interaction
between the adjacent arms. In the end, the asymptotic ($t\rightarrow \infty$)
form of the interacting component in the scattered state (\ref{eq:psiB2}) is
given by the expression
\begin{equation}
      \hat{S}\hat{f}_1^\dagger\hat{g}_2^\dagger\!=\! \int\! dxdy \,
      f(x)g(y)e^{-i\Phi_{12}(x,y)}
      \hat\Psi_1^\dagger(x)\hat\Psi^\dagger_2(y).
\end{equation}

Next, we calculate the two-particle density matrix~(\ref{eq:rhodef}) for the
scattering state~(\ref{eq:psiB2}) inside the interferometer, $\hat\rho_{BB'} =
\mbox{Tr}_{x,y}\bigl\{|BB^\prime\rangle \langle BB^\prime|\bigr\}$,
\begin{equation}
      \hat\rho_{BB'} = (\hat{s}_\alpha\otimes\hat{s}_{\alpha^\prime})
      \left[\begin{array}{cccc}
      1&1&1&{\cal V}\\
      1&1&1&{\cal V}\\
      1&1&1&{\cal V}\\
      {\cal V}^*&{\cal V}^*&{\cal V}^*&1
      \end{array}\right]
      (\hat{s}_\alpha\otimes\hat{s}_{\alpha^\prime}),
      \label{eq:rho2p}
\end{equation}
with $\hat{s}_\alpha = \mbox{diag}\{\sin\alpha,\cos\alpha\}$ and
\begin{equation}
      {\cal V} = \int dxdy\, |f(x)|^2 |g(y)|^2 \,
      e^{-i\Phi_{12}(x,y)}.
      \label{eq:visibility}
\end{equation}
We choose a particular form of the interaction kernels, $\kappa_1(x) =
\kappa_2(x) = \exp(-|x|/a)$, where $2a$ is the length of the interaction
region. Then the asymptotic form of the phase $\Phi_{12}(x,y)$ at $x,y
\rightarrow+\infty$ is given by,
\begin{equation}
      \Phi_{12}(x,y) =
      \varphi_0\exp\Bigl(-\frac{|x-y|}a\Bigr)\Bigl( 1
      +\frac{|x-y|}a \Bigr),
      \label{eq:intphase}
\end{equation}
with
\begin{equation}
      \varphi_0 = \frac{E_{12}\tau_0}{\hbar}
      \label{eq:phi0}
\end{equation}
the interaction-induced phase accumulated by the particles during the
simultaneous propagation through the interaction region; here, $\tau_0=a/v$
denotes the ballistic travelling time through the interaction region. We
consider the case of simultaneous injection and choose incoming wave-packets
of Lorentzian form with $\xi > 0$,
\begin{equation}
      f(x) = g(x) = \sqrt{\frac\xi\pi}\frac1{x+i\xi}.
      \label{eq:lorentzian}
\end{equation}
Then the parameter ${\cal V}$ can be expressed as a function of two
dimensionless parameters, the phase $\varphi_0$, see Eq.\ (\ref{eq:phi0}),
measuring the strength of the Coulomb interaction, and $\gamma = \xi/a$
quantifying the width of the incoming wave-packet,
\begin{equation}\label{eq:Nu}
   {\cal V}(\varphi_0,\gamma) = \!\int \frac{dx}{\pi} 
   \frac{\exp\bigl[ -i\varphi_0 e^{-2\gamma|x|}\bigl(1+2\gamma|x|\bigr) \bigr]}
   {x^2+1}.
\end{equation}

Consider first the situation with infinitely narrow incoming wave-packets
($\gamma\rightarrow 0$) simultaneously injected into the device. The relevant
values of the phase $\Phi_{12}(x,y)$ in ${\cal V}$ are those near $x \approx
y$, $\Phi_{12}(x,x) = \varphi_0$, resulting in 
\begin{equation}\label{eq:V}
   {\cal V}(\varphi_0,0) \rightarrow \exp(-i\varphi_0),
\end{equation}
see Eqs.\ (\ref{eq:visibility}) and (\ref{eq:intphase}) (when the wave-packets
are injected with a time delay $\tau_\mathrm{d}$, the phase $\varphi_c \sim
\Phi_{12}(v \tau_\mathrm{d}) \sim \varphi_0 \exp(-v\tau_\mathrm{d}/a)$ is
reduced).  The density matrix~(\ref{eq:rho2p}) corresponds to a pure state
with $\hat{\rho}_{BB'}^2 = \hat{\rho}_{BB'}$. According to
Eq.~(\ref{eq:horodecki}), the maximal value of the Bell parameter is given by,
\begin{equation}
      {\cal B}_\mathrm{max} = 2\sqrt{1 + \sin^2(2\alpha)
      \sin^2(2\alpha^\prime) \sin^2\frac{\varphi_0}2}.
      \label{eq:maxv2p}
\end{equation}
The maximal violation ${\cal B}_\mathrm{max} = 2\sqrt{2}$ is attained for
symmetric incoming beam splitters with $\alpha = \alpha^\prime = \pi/4$ and
$\varphi_0 = \pi$, requiring proper tuning of the interaction domain.

Next, we determine the optimal configurations of magnetic fluxes and angles
$(\Phi,\bar\Phi, \beta,\bar\beta)$ and $(\Phi^\prime, \bar{\Phi}^\prime,
\beta^\prime,{\bar\beta}^\prime)$ for the upper and lower interferometer where
the Bell inequality is maximally violated (we consider only the setup with
symmetric incoming beam splitters).  To do so, we derive the explicit form of
the Bell correlation function $E_{\beta\beta^\prime} (\Phi,\Phi^\prime)$, see
Eq.~(\ref{eq:bellcorr}). The correlation functions $\langle \hat{N}_i
\hat{N}_{i^\prime}\rangle$ entering into $E$ are given by the diagonal
elements of the two-particle density matrix $\hat\rho_\mathrm{out}$ after
scattering at the outgoing beam splitters, see Eq.~(\ref{eq:density_out}).
Then the Bell correlation function~(\ref{eq:bellcorr}) assumes the form (we
can set $\beta = \bar\beta$ and $\beta^\prime = {\bar\beta}^\prime$ and still
violate the Bell inequality maximally)
\begin{eqnarray}
      E &=& \frac12\sin2\beta\cos2\beta^\prime
      \bigl[V\cos(\Phi\! +\!\varphi_\mathrm{c}) - \cos\Phi\bigr]
      \label{eq:bellctwo}\\
      &+&\frac12 \cos2\beta\sin2\beta^\prime
      \bigl[V\cos(\Phi^\prime + \varphi_\mathrm{c})-\cos\Phi^\prime\bigr]
      \nonumber\\
      &-&\frac12\sin2\beta\sin2\beta^\prime
      \bigl[V\cos(\Phi\!+\!\Phi^\prime\!+\!\varphi_\mathrm{c})\!+\!
      \cos(\Phi\!-\!\Phi^\prime)\bigr],
      \nonumber
\end{eqnarray}
where $V = |{\cal V}|$ and $\varphi_\mathrm{c} = -\arg{\cal V}$. For
infinitely narrow wave-packets simultaneously injected into the device, we
have $V = 1$ and $\varphi_\mathrm{c} = \varphi_0$, see Eq.\ (\ref{eq:V}). The
Bell parameter then takes the form
\begin{eqnarray}
      &&{\cal B} = 2\bigl[\sin2\beta\cos2\beta^\prime
      \sin\Phi + \cos2\beta\sin2\beta^\prime
      \sin\Phi^\prime\bigr] \sin\frac{\varphi_\mathrm{c}}2
      \nonumber\\
      &&~~~~~~~+\sin2\beta\sin2\beta^\prime\bigl[ \cos\Phi\cos\Phi^\prime +
      \cos\bar\Phi\cos\Phi^\prime
      \label{eq:B2p}\\
      &&\qquad\qquad\qquad\qquad\qquad + \cos\Phi\cos\bar{\Phi}^\prime -
      \cos\bar\Phi\cos\bar{\Phi}^\prime\bigr], \nonumber
\end{eqnarray}
where we have shifted all fluxes by $\varphi_\mathrm{c}/2$, $\Phi \rightarrow
\Phi + \varphi_\mathrm{c}/2$ and so on. The maximal violation Eq.\
(\ref{eq:maxv2p}) is reached when
\begin{equation}
      (\Phi,\bar{\Phi}) \rightarrow \bigl(\pi/2,
      0\bigr), \quad
      (\Phi^\prime, \bar{\Phi}^\prime) \rightarrow \bigl(
      0, \pi\bigr)
      \label{eq:settings1}
\end{equation}
and with angles $\beta = \pi/4$ and $\beta^\prime = \mathrm{arcctg}[\sin(
\varphi_\mathrm{c}/2)]/2$ for the outgoing bean splitters.  Note, that one can
approach the maximal violation fixing the angles $\beta$ and $\beta^\prime$ of
the beam splitters to the above values and changing only the magnetic fluxes.
\begin{figure} 
   \includegraphics[width=7.0cm]{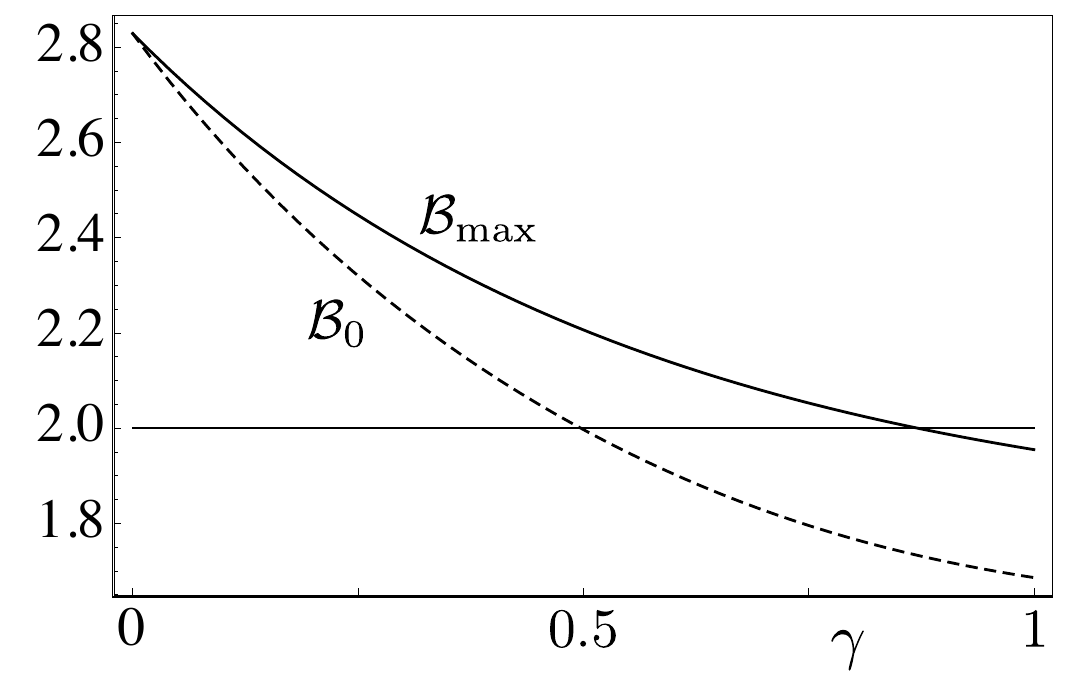}
   \caption[]
   {Bell parameters ${\cal B}_\mathrm{max}$ and ${\cal B}_0$ as a function of
   of width parameter $\gamma = \xi/a$ at $\varphi_0 = \pi$ for an optimal
   setting of Bell parameters (${\cal B}_\mathrm{max}$, solid line) as given
   by Eq.~(\ref{eq:horodecki}) and for the non-optimal settings in
   Eq.~(\ref{eq:settings1}) (${\cal B}_0$, dashed line).}
   \label{fig:bell1}
\end{figure}

In a next step, we consider the situation where the incoming wave-packets have
a finite width and thus a finite $\gamma$. Here, in contrast to the $\xi \to
0$ case, the interaction deforms the initial shape of the single particle
wave-functions.  This deformation appears only when both electrons are
transmitted through the interacting arms `1' and `2' and is absent in the
other scattering channels.  Thus the orbital entanglement in the scattering
(or pseudo-spin) degrees of freedom is recorded in the specific form of the
orbital wave-functions. Since our detection scheme captures only the total
number of the transmitted particles and is insensitive to the specific shape
of the wave-packets, the reduced density matrix $\hat\rho_{BB'}$ in Eq.\
(\ref{eq:rho2p}) is mixed and the pseudo-spin entanglement is reduced (for the
case of narrow wave-packets with $\xi \to 0$, the tracing over $x$ and $y$
generates a mixed state as well, but does not reduce the entanglement which is
only in the lead-indices).  A further reduction in entanglement of extended
wave packets is due to the electrons not meeting each other in the
finite-range interaction region, thereby reducing their effective interaction.
\begin{figure}
   \includegraphics[width=6.0cm]{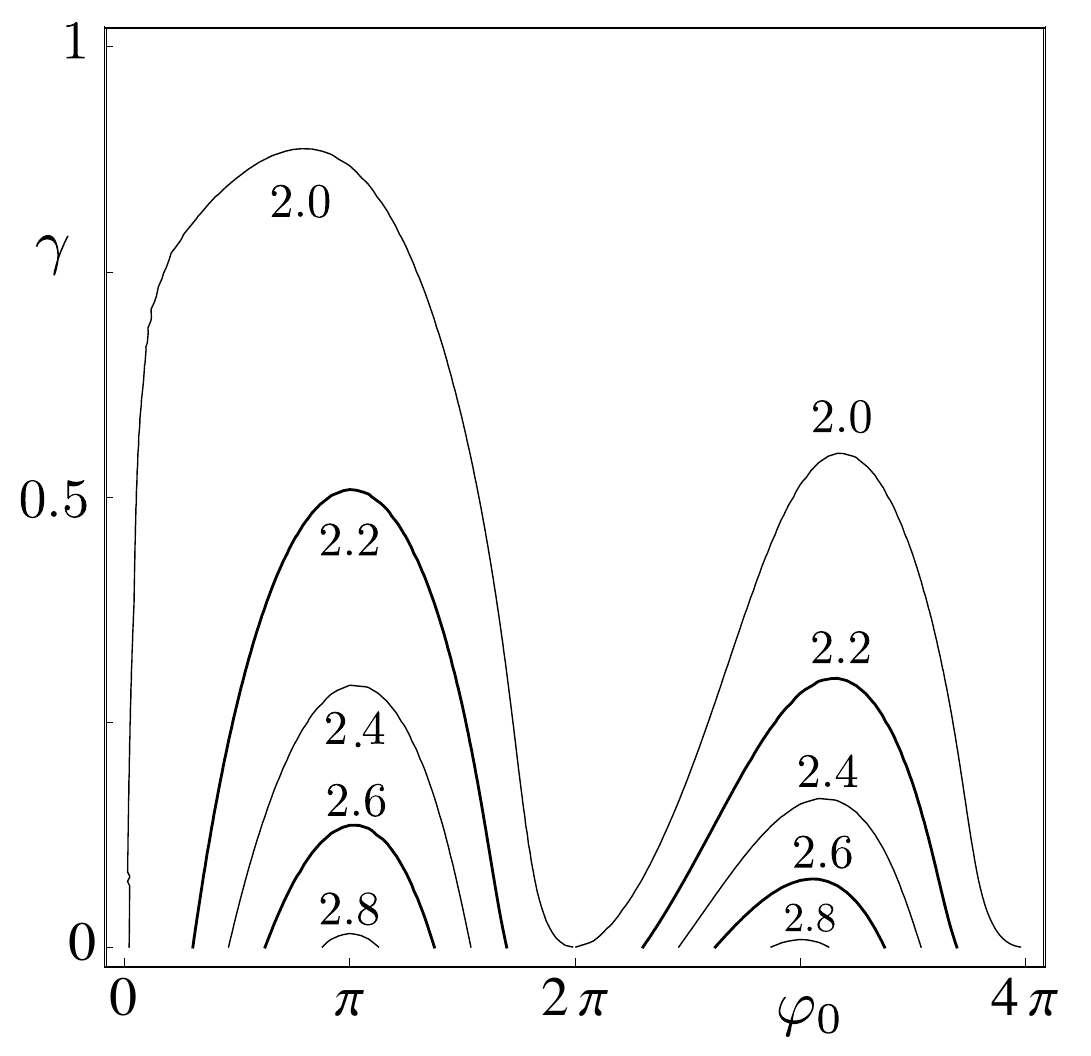}
   \caption[]
   {Contour plot of the Bell parameter ${\cal B}_\mathrm{max}$ for optimal
   settings of the Bell parameters as a function of the interaction parameter
   $\varphi_0$ (horizontal axis) and the width parameter $\gamma$ (vertical
   axis). For small-width, high-energy wave packets ($\gamma\to 0$) the
   Bell inequality is maximally violated for odd multiples of $\pi$.}
   \label{fig:bell2}
\end{figure}

When discussing the quantitative result for the Bell test with finite-width
wave packets, we first analyze the value of the Bell parameter for the
settings~(\ref{eq:settings1}) which have been optimized for the $\gamma = 0$
case,
\begin{equation}
      {\cal B}_0 = (1+V)\sqrt{1+\sin^2\frac{\varphi_\mathrm{c}}2}.
      \label{eq:bell2}
\end{equation}
In the most favorable situation where $\varphi_\mathrm{c} = \pi$, the Bell
inequality cannot be violated for $V < \sqrt{2} -1$. In reality, the
interaction phase $\varphi_\mathrm{c}$ monotonically decreases from $\pi$ as
$\gamma$ increases; in the limiting case of spatially extended wave-packets
$\gamma\rightarrow \infty$, the electrons are unlikely to meet each other in
the interaction region and $\varphi_\mathrm{c} \rightarrow 0$. 

The settings in Eq.\ (\ref{eq:settings1}), however, are not optimal when
$\gamma$ is finite. In Fig.\ \ref{fig:bell1}, we compare the maximal possible
value of the Bell parameter ${\cal B}_\mathrm{max}$ deriving from Eq.\
(\ref{eq:horodecki}) with the non-optimal value ${\cal B}_0$ as given by Eq.\
(\ref{eq:bell2}) choosing an interaction parameter $\varphi_0 = \pi$.  We find
that ${\cal B}_\mathrm{max} > {\cal B}_0$ and the Bell inequality can be
violated as long as $\gamma < \gamma_c \approx 0.87$ (the critical value for
${\cal B}_0$ is $\gamma_\mathrm{c} \approx 0.497$).

The maximal value of the Bell parameter in the entire parameter space
$(\varphi_0, \gamma)$ is shown in Fig.\ \ref{fig:bell2}. The region where the
Bell inequality is violated (${\cal B} > 2$) is shrinking as the strength of
the Coulomb interaction (as encoded in $\varphi_0$) grows. Indeed, for a finite
$\gamma$, a stronger interaction implies a larger deformation of the
wave-packets and thus more pseudo-spin entanglement is recorded in the shape
of the wave-packets. As a result, the detectable pseudo-spin entanglement is
reduced, thus lowering the Bell inequality violation. On the contrary, for
$\gamma = 0$, the maximal Bell parameter is a periodic function of $\varphi_0$
with maxima at $\varphi_0 = \pi, 3\pi,\,\dots$, see Eq.~(\ref{eq:maxv2p}).

\section{Bell experiment with Fermi Sea}\label{sec:BeFS}

We now extend our analysis to the more realistic situation including the (zero
temperature) Fermi sea. First, we study the scattering problem where a
single-particle Lorentzian wave-packet crosses an interaction region in a {\it
single} lead. This problem has been studied before in Ref.\
\onlinecite{lebedev11} and we restate the results here. Next, we consider the
scattering of two Lorentzian wave-packets propagating in two adjacent
interacting leads and make use of these results to find a many-electron
scattering state in the complete system with two coupled Mach-Zehnder
interferometers. Knowledge of this state then permits us to determine the
level of Bell inequality violation reachable in the presence of a Fermi sea.

\subsection{Scattering of a single wave-packet}

Consider a single-electron state with a wave-function $f(x)$ created
at $t_0 \rightarrow -\infty$ in the interaction-free asymptotic
region of a one-dimensional conductor,
\begin{equation}
      |f\rangle = \int dx\, f(x)
      \hat\Psi^\dagger(x,t_0)|\Phi_\mathrm{F}\rangle,
      \label{eq:single}
\end{equation}
where $|\Phi_\mathrm{F}\rangle$ is the zero-temperature Fermi sea.
We assume that all Fourier components of $f(x) = \sum_k f_k
e^{ikx}$ are vanishing below the Fermi momentum, $f_k = 0$ for $k <
k_\mathrm{F}$. For the specific case of a Lorentzian wave packet,
see Eq.~(\ref{eq:lorentzian}), the incoming state $|f\rangle$ can be
created by applying a local voltage pulse of Lorentzian shape,
\begin{equation}
      V(t) = \frac{2\hbar}{e\tau}\, \frac{1}{1 +
      (t-t_0)^2/\tau^2},
      \label{eq:lpulse}
\end{equation}
in the asymptotic region $x_0 \rightarrow -\infty$ of the conductor; here
$\tau  = \xi/v_\mathrm{F}$ is the duration of the pulse\cite{driftvel}. Then
the state $|f\rangle$ can be written as\cite{delft},
\begin{equation}
      |f\rangle = \hat{U}[\phi(t)] \hat{F}^\dagger|\Phi_\mathrm{F}\rangle,
      \label{eq:pulsestate}
\end{equation}
where $\hat{F}^\dagger$ is an electron ladder operator increasing the number
of electrons in the system by one. The unitary operator $\hat{U}[\phi]$
describes the evolution of the wave function under the action of the
Lorentzian voltage pulse~(\ref{eq:lpulse}) applied at the position $x = x_0$,
\begin{equation}
      \hat{U}[\phi] = \hat{T}_+\exp\Bigl[ iv_\mathrm{F}\! \int dt\,
      \phi(t)\hat\rho(x_0,t) \Bigr],
      \label{eq:Udef}
\end{equation}
with $\hat{T}_+$ the forward time-ordering operator, $\hat\rho(x,t)
=$ ${:\hat\Psi^\dagger(x,t) \hat\Psi(x,t):}$ the electron density operator, and
$\phi(t) = \int^t dt^\prime eV(t^\prime)/\hbar$ is the phase accumulated by
the electrons.

The wave packet then propagates ballistically towards the interaction region
around $x = 0$ and arrives there at $t = 0$, assuming $x_0 = v_\mathrm{F}t_0$.
Due to the interaction, the excess electron can exchange energy with the Fermi
sea, thereby exciting electron-hole pairs.  Subsequently, this many-particle
scattering state (the excess electron plus the electron-hole cloud)
ballistically propagates to the interaction-free region at large $x
\rightarrow +\infty$ where it assumes the form
\begin{equation}
      |\tilde{f}\rangle = \hat{T}_+\exp\Bigl[
      -\frac{i}\hbar\int_{-\infty}^\infty dt \, \hat{H}_\mathrm{int}(t)
      \Bigr]|f\rangle,
\end{equation}
with $\hat{H}_\mathrm{int}$ given in Eq.~(\ref{eq:hint}).

At first glance the scattering state $|\tilde{f}\rangle$ should be a rather
complicated entangled state involving an excess electron and additional
electron-hole pairs.  However, the central result of Ref.\
\onlinecite{lebedev11} says that the state $|\tilde{f}\rangle$ can again be
produced by applying an additional voltage pulse $V_\chi(t)$ to the incoming
state $|f\rangle$, $|\tilde{f}\rangle = \hat{U}[\chi]|f\rangle$, with
$\chi(t) = \int^t dt^\prime\, e\, V_\chi(t^\prime)/\hbar$. The remarkable
consequence of this fact is that the scattering state has the form of a simple
Slater determinant and is in fact non-entangled, since it can be produced by
an evolution operator generated by a single particle Hamiltonian.

The peculiarity of the Lorentzian incoming state becomes clear when one goes
to the bosonic formulation, expressing the fermionic field operator through
bosonic creation/annihilation operators $\hat{b}_k^\dagger$ and $\hat{b}_k$,
$\hat\Psi(x) \propto \exp[i\sum_{k>0} (\hat{b}_k e^{ikx} + \hat{b}_k^\dagger
e^{-ikx})/\sqrt{k}] \hat{F}$, see Appendix for details. Integrating over the
coordinate in the definition of the incoming state (see Eq.~(\ref{eq:single}))
with the Lorentzian form of $f(x)$, the wave-packet can be written in the form
\begin{equation}
      |f\rangle \propto \exp\Bigl( -i
      \sum_{k>0} \frac{e^{-k\xi}}{\sqrt{k}}\, \hat{b}_k^\dagger \Bigr)
      \hat{F}^\dagger |\Phi_\mathrm{B}\rangle,
\end{equation}
where $|\Phi_\mathrm{B}\rangle$ is the bosonic vacuum state. This state can be
recognized as a coherent state of bosons  $|f\rangle = \prod_{k>0}|v_k\rangle$
with amplitudes $v_k = -ie^{-k\xi}/\sqrt{k}$. In the bosonized picture any
electron interaction of the form $V(x,y)\hat\rho(x)\hat\rho(y)$ is quadratic
in bosonic variables and thus corresponds to a potential scattering of the
bosons. Due to the chiral nature of the present scattering problem, the back
reflection of bosons is forbidden and the only result of that scattering is
the appearance of a momentum dependent forward scattering phase
$\delta_\mathrm{sc}(k)$, $\hat{b}_k \rightarrow \hat{b}_k \exp[i
\delta_\mathrm{sc}(k)]$, where the actual form of $\delta_\mathrm{sc}(k)$
depends on the particular form of the interaction kernel $V(x,y)$. Hence, the
scattered state of bosons is again a {\it coherent} state,
\begin{equation}
      |\tilde{f}\rangle \propto \exp\Bigl[
      -i\sum_{k>0} \frac{e^{-k\xi -i\delta_\mathrm{sc}(k)}}{\sqrt{k}}\, 
      \hat{b}_k^\dagger \Bigr] \hat{F}^\dagger |\Phi_\mathrm{B}\rangle.
      \label{eq:coherent}
\end{equation}
Going back to the electronic picture, this scattered state can again be
created by applying a corresponding voltage pulse $\tilde{V}(t)$ given by
\begin{equation}
      \tilde{V}(t) = \frac\hbar{e} \int\limits_0^\infty d\omega\,
      \exp[-\tau \omega +i \delta_\mathrm{sc}(\omega/v_\mathrm{F}) +i\omega
      t] + C.c.
\end{equation}
Even more, for an initial voltage pulse of arbitrary form, the generated state
$|f_U\rangle = \hat{U}[\phi] \hat{F}^\dagger |\Phi_\mathrm{F}\rangle$ can be
represented as a coherent bosonic state as well, $|f_U\rangle = \prod_{k>0}
e^{u_k \hat{b}_k - u_k^* \hat{b}_k^\dagger} \hat{F}^\dagger
|\Phi_\mathrm{B}\rangle$ with amplitudes $u_k = v_\mathrm{F}\sqrt{k}\,
\phi(kv_\mathrm{F})/2\pi$ determined by the Fourier transform $\phi(\omega)$
of the phase $\phi(t) = \int^t dt^\prime eV(t^\prime)/\hbar$. The voltage
pulse $\tilde{V}(t)$ generating the wave function behind the scatterer then
derives from the phase $\tilde\phi(\omega) = \phi(\omega)
\exp[i\delta_\mathrm{sc}(\omega/ v_\mathrm{F})]$.

\subsection{Scattering of two wave-packets}

Let us next analyse the scattering problem for two electron wave-packets
propagating in the two adjacent interacting leads (`1' and `2') of the upper
and lower interferometer (unprimed and primed). We assume that initially at
$t_0 \rightarrow -\infty$ the incoming state is given by,
\begin{equation}
      |fg\rangle = \!\int\! dx dx^\prime f(x) g(x^\prime)
      \hat\Psi_1^\dagger(x,t_0) \hat\Psi^\dagger_2(x^\prime,t_0)
      |\Phi_\mathrm{F}\rangle.
\end{equation}
In order to generalize the result of the previous section to the case of the
two-particle scattering problem, we calculate the overlap of the scattered
state $|\tilde{f} \tilde{g}\rangle = \hat{S} |fg\rangle$ and the state
$|fg\rangle_{U}$ (for later convenience we use $\chi$ and $\chi'$ referring to
the upper and lower interferometers rather than the lead indices `1' and `2'),
\begin{equation}
      |fg\rangle_{U} = \hat{U}_1[\chi]
      \hat{U}_2[\chi']|fg\rangle,
\end{equation}
obtained by applying two distinct voltage pulses $V_{\chi}(t)$ and
$V_{\chi'}(t)$ in the corresponding (upper and lower) leads to the incoming
state $|fg\rangle$ in the {\it non-interacting} system. The overlap,
\begin{equation}
      {\cal O}_{fg}[\chi(t),\chi'(t)] = \langle
      fg|\hat{U}_1^\dagger[\chi] \hat{U}^\dagger_2[\chi']
      \hat{S}|fg\rangle,
      \label{eq:overlap}
\end{equation}
is a function of the phases $\chi(t)$ and $\chi'(t)$ and we want to find its
maxima, ${\cal O}_{fg} = \max_{\chi,\chi'}|{\cal O}[\chi,\chi']|$ as a
function of phases $\chi(t)$ and $\chi'(t)$.

As shown in the Appendix, there exist particular phases $\chi(t)$ and
$\chi'(t)$ producing an overlap~(\ref{eq:overlap}) of unit magnitude,
provided that the incoming wave-packets $f(x)$ and $g(x)$ have a Lorentzian
form (the more general statement actually is that one can reach full overlap
for {\it any} voltage-pulse generated incoming states). Thus we conclude that
the scattered state $|\tilde{f}\tilde{g}\rangle = \hat{S}|fg\rangle$ can be
obtained by applying two voltage pulses to the incoming state of the
non-interacting problem,
\begin{equation}
      \hat{S}|fg\rangle = \exp(i\Phi) \hat{U}_1[\chi(t)]
      \hat{U}_2[\chi'(t)]|fg\rangle
\end{equation}
with $\Phi$ an overall phase.  As a result, the scattered state is a Slater
determinant and thus is non-entangled.

In the following, we assume for simplicity that the interacting leads are
characterized by the same coordinate kernels $\kappa_1(x) = \kappa_2(x) =
\kappa(x)$. Let $\vec{e}_\nu = \{e_{\nu1},e_{\nu2}\}$ and $\epsilon_\nu$, $\nu
= 1,2$, be the eigenvectors and corresponding eigenvalues of the coupling
matrix of the interaction Hamiltonian Eq.~(\ref{eq:hint}). Provided the
wave-packets $f(x)$ and $g(x)$ are Lorentzians with widths $\xi_1$ and $\xi_2$,
the interaction equivalent phases $\chi(t)$ and $\chi'(t)$ have Fourier
components of the form,
\begin{widetext}
\begin{eqnarray}
      &&\chi(\omega>0) = e_{11}(e_{11}e^{-\omega\tau_1} +
      e_{12}e^{-\omega\tau_2}) F_1(\omega) +
      e_{21}(e_{21}e^{-\omega\tau_1} +
      e_{22}e^{-\omega\tau_2})F_2(\omega),
      \label{eq:phiup}
      \\
      &&\chi'(\omega>0)= e_{12}(e_{11}e^{-\omega\tau_1} +
      e_{12}e^{-\omega\tau_2}) F_1(\omega) +
      e_{22}(e_{21}e^{-\omega\tau_1} +
      e_{22}e^{-\omega\tau_2})F_2(\omega),
      \label{eq:phidn}
\end{eqnarray}
with $\chi(-\omega) = \chi^*(\omega)$ and the same for $\chi'(\omega)$ and the
overall phase reads
\begin{eqnarray}
      \Phi = -\sum_{i=1,2}\int_0^\infty
      \frac{d\omega}{2\pi}\Bigl[ v_\mathrm{F}^2 \mbox{Re}\,[G_{++}(\omega)] 
      |F_i(\omega)|^2
      + \mbox{Re}\,[F_i(\omega)]\Bigr]
      \bigl( e_{i1}e^{-\omega\tau_1} + e_{i 2}
      e^{-\omega\tau_2} \bigr)^2,
      \label{eq:phase}
\end{eqnarray}
\end{widetext}
where $\tau_{1,2} = \xi_{1,2}/v_\mathrm{F}$. The functions $F_i(\omega)$ are
given by
\begin{equation}
      F_i(\omega) =
      \frac{\epsilon_i|\kappa(\omega)|^2}{1-i\epsilon_i\Pi_{++}^*(\omega)},
      \label{eq:F}
\end{equation}
with $\kappa(\omega) = \int dt\, \kappa(v_\mathrm{F}t) e^{-i\omega t}$,
$\Pi_{++}(\omega) = \int dxdx^\prime \kappa(x)$ $\kappa(x^\prime)\,
G_{++}(\omega,x,x^\prime)$ and $G_{++}(\omega,x,x^\prime)$ is the Fourier
transform of the Green's function $G_{++}(\tau,x,x^\prime) =
\langle\hat{T}_+\{\hat\rho(x,\tau) \hat\rho(x^\prime,0)\}\rangle$,
\begin{equation}
      G_{++}(\omega,x,x^\prime) =\frac1{2\pi v_\mathrm{F}^2}\int
      \frac{d\omega^\prime}{2\pi i}\frac{\omega^\prime
      e^{-i\omega^\prime(x-x^\prime)/v_\mathrm{F}}}{\omega^\prime - \omega
      - i\delta \mathrm{sgn}(\omega)}.
      \label{eq:rr}
\end{equation}

\subsection{The complete scattering state}

Due to the self-capacitance of the interferometer arms, electron-hole pairs
can be excited in scattering processes where only one wave-packet (the upper
or lower) propagates through the corresponding interacting arm of the
interferometer. The corresponding interaction equivalent fields $\chi$ and
$\chi'$ and the overall phase $\Phi$ can be formally found from
Eqs.~(\ref{eq:phiup})--(\ref{eq:phase}) by setting the width of the absent
wave-packet to infinity, $\xi_1 \rightarrow \infty$ or $\xi_2 \rightarrow
\infty$. In the following, we introduce a new notation for the interaction
equivalent fields in the upper ($\chi_{ij}$) and lower ($\chi'_{ij}$)
interferometer and for the overall phase ($\Phi_{ij}$), where the indices $i,j
\in \{0,1\}$ tell whether a particle has moved through the upper ($i=1$) or
lower ($j=1$) lead (else $i,j = 0$ if there is no particle in the lead).  The
scattering state inside the interferometers before the outgoing beam splitters
$B$ and $B'$ then takes the form (we assume symmetric incoming beam splitters with
$\alpha=\alpha^\prime = \pi/4$; note that the operators $\hat U$ always act on
the interacting leads `1' and `2'),
\begin{eqnarray}
      |BB'\rangle &=& \frac12\bigl[
      \hat{f}^\dagger_3 \hat{g}^\dagger_4
      + e^{i\Phi_{11}}
      \hat{U}_1[\chi_{11}] \hat{U}_2[\chi_{11}^\prime]
      \hat{f}^\dagger_1 \hat{g}^\dagger_2
      \nonumber\\
      &&\quad
      + e^{i\Phi_{10}} \hat{U}_1[\chi_{10}]
      \hat{U}_2[\chi^\prime_{10}] \hat{f}^\dagger_1
      \hat{g}^\dagger_4
      \nonumber\\
      &&\quad +e^{i\Phi_{01}}
      \hat{U}_1[\chi_{01}] \hat{U}_2[\chi_{01}^\prime]
      \hat{f}^\dagger_3 \hat{g}^\dagger_2 \bigr] |\Phi_\mathrm{F}\rangle.
      \label{eq:psiBF}
\end{eqnarray}
Each of the four scattered- or pseudo-spin components corresponds to a Slater
determinant state and thus does not exhibit any interaction-induced
entanglement. However, the overall state cannot be factorized with respect to
the upper and lower interferometer degrees of freedom because each component
is formally obtained under a different evolution (different
interaction-equivalent phases). Hence, the interaction induces pseudo-spin
entanglement between the non-entangled many-particle electron-hole cloud
states.

Before evaluating the Bell inequalities for the full setup including the Fermi
sea, we first wish to gauge our expectations with a discussion of the degree
of entanglement we can expect in our two Mach-Zehnder interferometers.  Our
Bell test is sensitive only to the excess particle number transmitted through
a given outgoing lead. In order to describe the statistics of the Bell test,
it is sufficient to know the reduced $4 \times 4$ density matrix
$\hat\rho_{BB'}$ obtained from the {\it full} density operator
$\hat\rho_\mathrm{ud} = |BB' \rangle \langle BB'|$ by tracing out all spatial
degrees of freedom, see Eq.\ (\ref{eq:rhodef}). Formally, this corresponds to
dividing all degrees of freedom into the (obervable) pseudo-spin degrees of
freedom associated with finding a particle in the leads `1' or `3' (encoded by
$B$) or in the leads `2' or `4' (encoded in $B'$) plus all the remaining
degrees of freedom (we encode these with the letter $R$), hence
$\hat\rho_{BB'} = \mbox{Tr}_R  |BB' \rangle \langle BB'|$. In a first step, we
then can calculate the von Neumann entropy $S(\hat\rho_{BB'}) \equiv -
\mbox{Tr}\bigl[\hat\rho_{BB'} \log_2(\hat\rho_{BB'}) \bigr]$ of the measured
outcomes.  This entropy ranges between 0 and 2 and tells us about the degree
of entanglement between the pseudo-spin degrees of freedom and the rest of the
system. A value $S(\hat\rho_{BB'}) = 0$ tells us, that $R$ is not entangled
with the pseudo-spin system $BB'$ and hence the system $BB'$ can be maximally
entangled; on the contrary, a value $S(\hat\rho_{BB'}) = 2$ informs us that
$R$ is fully entangled with $BB'$ and all information about the latter is
encoded in $R$. Due to the monogamy of entanglement, the two subsystems $B$
and $B'$ cannot be entangled in this case.  Indeed, calculating the reduced
densitiy matrices $\hat\rho_{B} = \mbox{Tr}_{B'}[ \rho_{BB'}]$ and
$\hat\rho_{B'} = \mbox{Tr}_{B} [\rho_{BB'}]$ and the associated entropy
$S(\hat\rho_{B}) \equiv - \mbox{Tr} \bigl[ \hat\rho_{B}
\log_2(\hat\rho_{B}) \bigr]$ (and similar for $B'$) we expect that
$S(\hat\rho_{B})$ ranges between 0 ($B$ not entangled with $B'$) and 1 ($B$
maximally entangled with $B'$) if $S(\hat\rho_{BB'}) = 0$, while
$S(\hat\rho_{B}) =0$ if $S(\hat\rho_{BB'}) = 2$.
\begin{figure}
   \includegraphics[width=6.0cm]{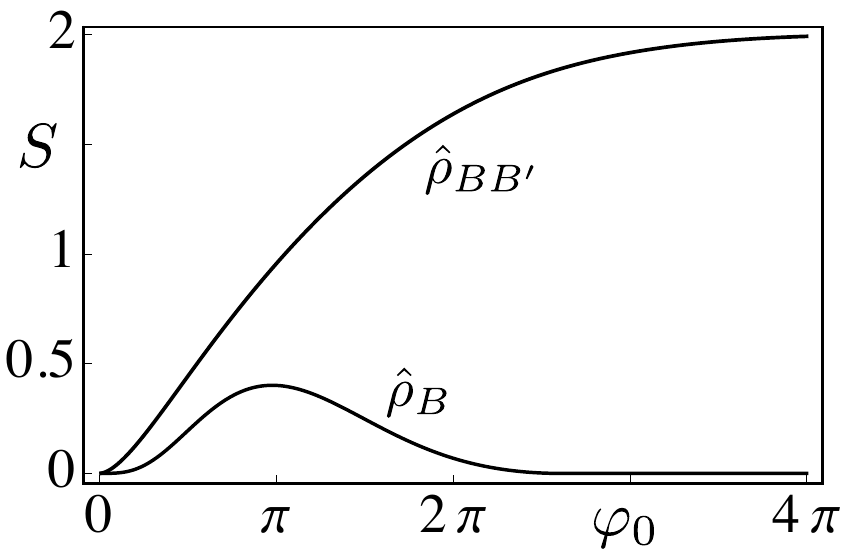}
   \caption[]
   {\label{fig:S} Von Neumann entropy $S(\hat{\rho}_{BB'})$ of measured
   outcomes of the double Mach-Zehnder interferometer quantifying the
   entanglement with the unobserved degrees of freedom in the system. The
   entanglement increases with the interaction strength and the system is
   fully entangled with the environment at large $\varphi_0$. The entropy
   $S(\hat{\rho}_B)$ expresses the entanglement {\it between} the two
   Mach-Zehnder interferometers and thus the potential for violation of the
   Bell test.  Increasing the interaction first entangles the two MZ
   interferometers with each other but as the entire system gets entangled
   with the environment upon further increase of $\varphi_0$, the entanglement
   between the interferometers decreases.}
\end{figure}

Hence, the more degrees of freedom from $R$ that are entangled with $B$ or
$B'$ we integrate out (as we do not observe them in our Bell test), the
smaller is the remaining entanglement between $B$ and $B'$ and the weaker is
the violation of the Bell test, see Fig.\ \ref{fig:S}.  Consider first the
two-particle Bell test described in Sec.\ \ref{sec:2pBt}. For narrow wave
packets, the trace over the coordinates does not reduce the entanglement in
the outgoing leads (hence $S(\hat\rho_{BB'})=0$) and we expect the entropy
$S(\hat\rho_B)$ to reach unity under ideal conditions where $B$ and $B'$ are
fully entangled; we then can expect that the Bell inequality can be maximally
violated. For extended wave packets, the spacial degrees of freedom are
relevant, e.g., some `which-path' information is stored in the shape of the
wave-function in the outgoing channels. The entropy $S(\hat\rho_{BB'})$ does
not vanish as the system $BB'$ is entangled with $R$ and hence the entropy
$S(\hat\rho_B)$ cannot reach unity any more; the Bell inequality cannot be
maximally violated any longer. Finally, including the Fermi sea, the
entanglement between $R$ and $BB'$ is even larger since even further
information on the $BB'$ system is encoded in unobserved degrees of freedom in
$R$, e.g., the presence of an additional hole in the outgoing channel tells,
that the particles sure did not choose the paths `3' and `4', since no
interaction is active in this case.  As a consequence, $S(\hat\rho_{BB'})$
deviates more strongly from zero, $S(\hat\rho_B)$ is further reduced from
unity, and the maximum in the Bell inequality violation is further diminished.

For a quantitative analysis, we then calculate the two-particle density matrix
$\hat\rho_{BB'}$ corresponding to the state~(\ref{eq:psiBF}), see
Eq.~(\ref{eq:rhodef}). The diagonal elements are all equal to $1/4$, while the
remaining six independent elements are given by
\begin{eqnarray}
      {[\hat\rho_{BB'}]}_{12,14} &=& \frac{-1}{4\pi}\int 
      \frac{\tau_2 dt}{\tau_2^2+t^2}\,
      e^{i[\chi^\prime_{11}]_+(t) + i[\chi_{10}^\prime]_-(t)},
      \label{eq:r1110}
      \\
      {[\hat\rho_{BB'}]}_{12,32} &=& \frac{-1}{4\pi}\int \frac{\tau_1
      dt}{\tau_1^2+t^2}\, e^{i[\chi_{11}]_+(t) +i[\chi_{01}]_-(t)},
      \label{eq:r1101}
      \\
      {[\hat\rho_{BB'}]}_{12,34} &=& \frac14\exp\bigl(
      i[\chi_{11}]_+(i\tau_1) + i[\chi^\prime_{11}]_+(i\tau_2) \bigr),~~~~~
      \label{eq:r1100}
\end{eqnarray}
\begin{eqnarray}
      {[\hat\rho_{BB'}]}_{14,34} &=&
      -\frac14\exp\bigl(i[\chi_{10}]_+(i\tau_1) \bigr),
      \label{eq:r1000}
      \\
      {[\hat\rho_{BB'}]}_{32,34} &=& -\frac14 \exp\bigl(
      i[\chi_{01}^\prime]_+(i\tau_2) \bigr),
      \label{eq:r0100}
      \\
      {[\hat\rho_{BB'}]}_{14,32} &=& \frac14\Bigl[ \frac1\pi
      \int\frac{\tau_1  dt}{\tau_1^2+t^2}\, e^{i[\chi_{10}]_+(t)
      + i[\chi_{01}]_-(t)} \Bigr]
      \label{eq:r1001}
      \\
      &&\quad \times \Bigl[ \frac1\pi\int\frac{\tau_2
      dt}{\tau_2^2+t^2}\, e^{-i[\chi_{01}^\prime]_-(t) -
      i[\chi_{10}^\prime]_+(t)} \Bigr],
      \nonumber
\end{eqnarray}
where $[f]_\pm(t)$ are the components of the function $f(t)$
analytical in the upper (lower) half plane,
\begin{equation}
      [f]_\pm(t) = \pm\frac1{2\pi i} \int dt^\prime\,
      \frac{f(t^\prime)}{t^\prime - t \mp i0 }.
\end{equation}
For the sake of simplicity we assume a symmetric situation where the
interacting leads have the same self-interaction $E_{11} = E_{22}$ and the
wave-packets have the same form $f(x) = g(x)$ and thus $\tau_1=\tau_2=\tau$,
implying that ${[\hat\rho_{BB'}]}_{12,14} = {[\hat\rho_{BB'}]}_{12,32}$ and
${[\hat\rho_{BB'}]}_{14,34} = {[\hat\rho_{BB'}]}_{32,34}$.  We first focus on
the situation of infinitely narrow wave-packets $\gamma =\xi/a \rightarrow
0$, allowing us to replace $\tau/(\tau^2+t^2) \rightarrow \pi\delta(t)$ in
Eqs.\ (\ref{eq:r1110})--(\ref{eq:r1001}). The Bell correlation function then
takes the form
\begin{eqnarray}
      E \!&=&\! \frac{V_1}2 \sin2\beta
      \cos2\beta^\prime\bigl[\cos(\Phi\! +\! \varphi_s\! +\!
      \varphi_\mathrm{c}) - \cos(\Phi\! +\!
      \varphi_\mathrm{s})\bigr]
      \nonumber\\
      &+&\!\frac{V_1}2 \cos2\beta
      \sin2\beta^\prime\bigl[\cos(\Phi^\prime\! +\! \varphi_s\! +\!
      \varphi_\mathrm{c}) - \cos(\Phi^\prime\! +\!
      \varphi_\mathrm{s})\bigr]
      \nonumber\\
      &-&\!\frac{1}{2}\sin2\beta \sin2\beta^\prime
      \bigl[(V_1V_2)^2\cos( \Phi\! +\! \Phi^\prime\! +\! 2\varphi_\mathrm{s}
      \!+\! \varphi_\mathrm{c})
      \nonumber\\
      &&\quad\qquad\qquad\qquad+ (V_1/V_2)^2\cos( \Phi\! -\!
      \Phi^\prime) \bigr]. 
      \label{eq:bellcfermi}
\end{eqnarray}
Here, the phase $\varphi_c = \mbox{Re}[\chi_{01}]_+(t=0)$ assumes the role of
$\varphi_c$ in the two-particle case, the phase difference between the two
cases where a particle traverses the lead `1' with ($\chi_{11}$) and without
($\chi_{10}$) presence of a particle in lead `2'.  Since $\chi_{11} =
\chi_{10}+\chi_{01}$, it is the difference $\chi_{01}$ that enters into
$\varphi_c$. The phase $\varphi_s = \mbox{Re}[\chi_{10}]_+(t=0)$ is due to the
self-interaction of the particle passing through lead `1'; this phase is
absent in the two-particle scattering problem. Finally, the visibility factors
are given by $V_1 = \exp\bigl[-\mbox{Im}[\chi_{10}]_+(t=0)\bigr]$ (describing
the situation where only one particle traverses the interaction region) and
$V_2 = \exp\bigl[-\mbox{Im}[\chi_{01}]_+(t=0)\bigr]$ (showing up when both
particles traverse the interaction region; no correlations appear in the
absence of any interaction). Note that both factors $V_1 V_2 <1$ and $V_1/V_2
< 1$.
\begin{figure}
   \includegraphics[width=6.0cm]{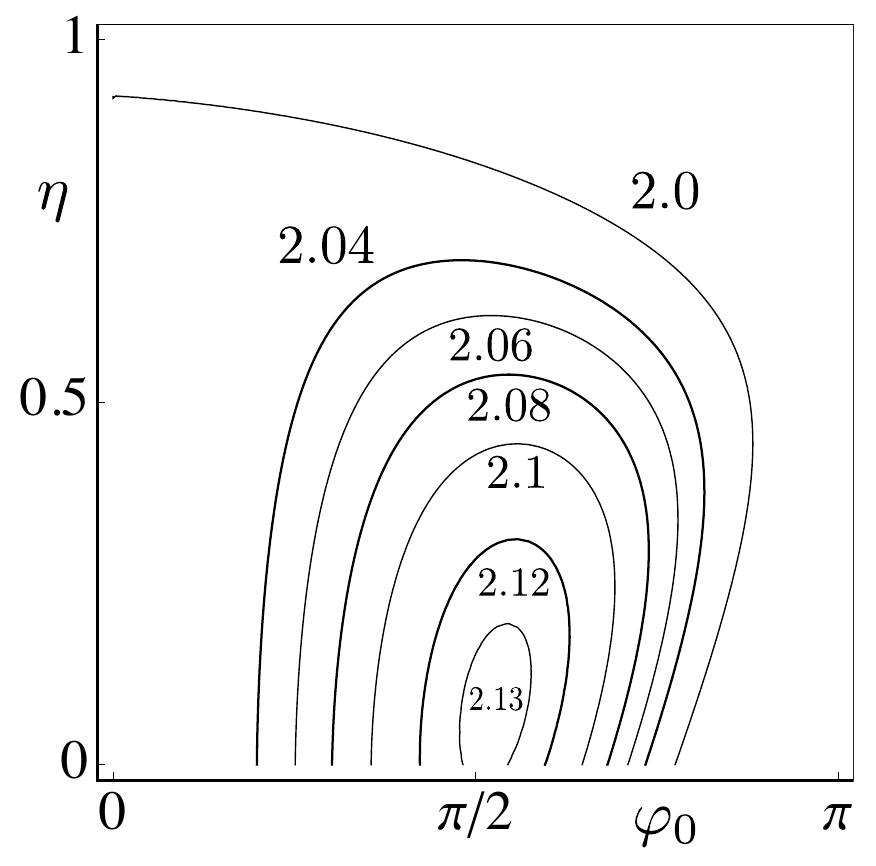}
   \caption[]
   {Bell parameter~(\ref{eq:bellfermi}) for the non-optimal settings of the
   Bell parameters in Eq.\ (\ref{eq:settings1}) as a function of interaction
   parameter $\varphi_0$ (horizontal axis) and the parameter $\eta$ (vertical
   axis) describing the ratio of self- to mutual capacitance.}
   \label{fig:bell3}
\end{figure}

The phase shift $\varphi_\mathrm{s}$ can be eliminated from Eq.\
(\ref{eq:bellcfermi}) by incorporation in the magnetic fluxes and setting
$\varphi_s = 0$.  Let us then choose the same values of (renormalized)
magnetic fluxes as for the two-particle case, Eq.\ (\ref{eq:settings1});
substituting the settings in Eq.\ (\ref{eq:settings1}) into Eq.\
(\ref{eq:bellcfermi}), we find the Bell parameter
\begin{equation}
      {\cal B} = 2V_1\sqrt{V_1^2 \frac{(V_2^2+V_2^{-2})^2}4 +
      \sin^2\frac{\varphi_\mathrm{c}}2},
      \label{eq:bellfermi}
\end{equation}
at $\beta = \pi/4$ and $\cot2\beta^\prime = 2\sin(\varphi_\mathrm{c}/2) /
(V_1(V_2^2 + V_2^{-2}))$.  As before, we consider the interaction kernel
$\kappa(x) = \exp(-|x|/a)$ and relate the self-coupling constants $E_{ii}$ to
the mutual coupling parameter $E_{12}$ via the dimensionless factor $\eta >
0$, $E_{11} = E_{22} = \eta E_{12}$. Then
\begin{equation}
      \kappa(\omega) = \frac{2\tau_0}{1+(\omega\tau_0)^2},
\end{equation}
and 
\begin{equation}
      \Pi_{++}(\omega) = -\frac{\tau_0}{2\pi i}\,
      \frac1{(\omega\tau_0+i)^2},
\end{equation}
with $\tau_0 = a/v_\mathrm{F}$ the ballistic traveling time through
the interaction region. For the symmetric setup, the interaction
equivalent phases obey the symmetry relations $\chi_{10} =
\chi_{01}^\prime$ and $\chi_{01} = \chi_{10}^\prime$ with
\begin{equation}
      \chi_{10}(t) = 4n_{10}\int d\nu\, \frac{e^{-|\nu|\gamma + i\nu
      t/\tau_0}}{(\nu+i)^2[(\nu-i)^2-n_{10}]},
      \label{eq:phsym}
\end{equation}
where $n_{10} = (\varphi_0/2\pi)(\eta + 1)$, and similarly for the $\chi_{01}$
field with $n_{01} = (\varphi_0/2\pi)(\eta - 1)$.  In Fig.\ \ref{fig:bell3} we
plot the value of the Bell parameter (\ref{eq:bellfermi}) as a function of
$\varphi_0$, see Eq.\ (\ref{eq:phi0}), and the parameter $\eta$. The maximal
value ${\cal B} \approx 2.13$ violating the Bell inequality is assumed at
$\varphi_0 \approx 0.53\, \pi$ and $\eta = 0.083$.

\begin{figure}
   \includegraphics[width=6.0cm]{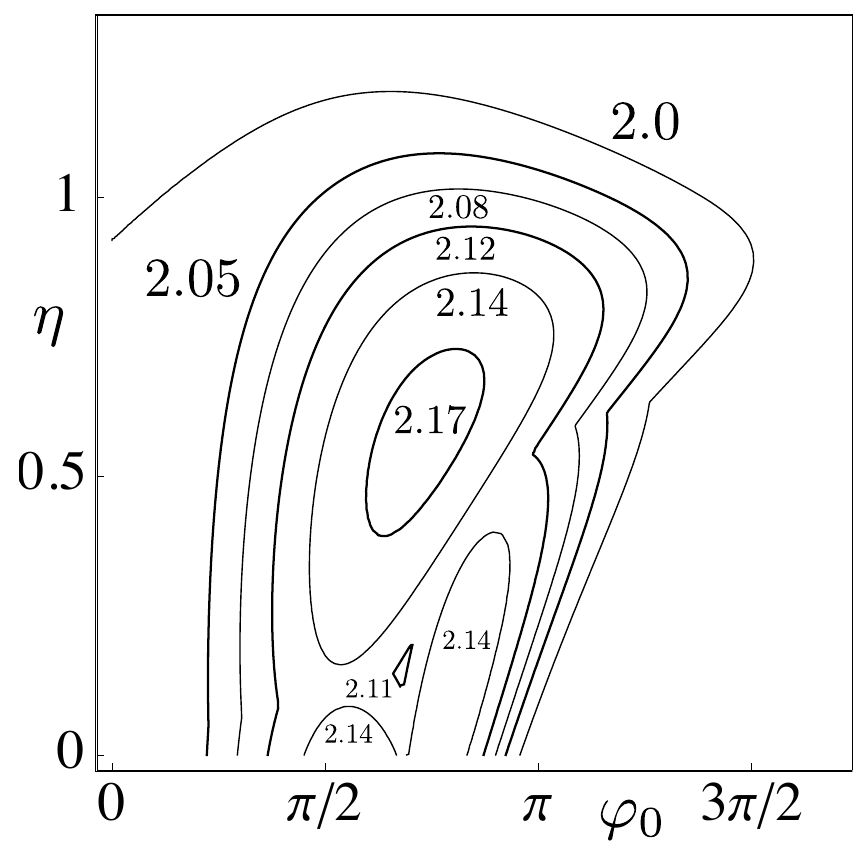}
   \caption[]
   {Maximal value of the Bell parameter as a function of interaction parameter
   $\varphi_0$ (horizontal axis) and the parameter $\eta$ describing the ratio
   of self- to mutual capacitance (vertical axis).}
   \label{fig:bell4}
\end{figure}

As noticed before, the settings (\ref{eq:settings1}) might be non-optimal in
the general case. The full optimization of the Bell test according to Eq.\
(\ref{eq:horodecki}) gives a higher value of the Bell parameter, see Fig.\
\ref{fig:bell4}, where we find the maximal violation of the Bell inequality
${\cal B} \approx 2.18 > 2$ at $\varphi_0 \approx 0.73\, \pi$ and $\eta \approx
0.58$. Note that a value $\eta < 1$ implies that the self-capacitance is
smaller than the mutual capacitance, $E_{22} < E_{12}$; we will discuss this
point further below.

\begin{figure}
   \includegraphics[width=7.0cm]{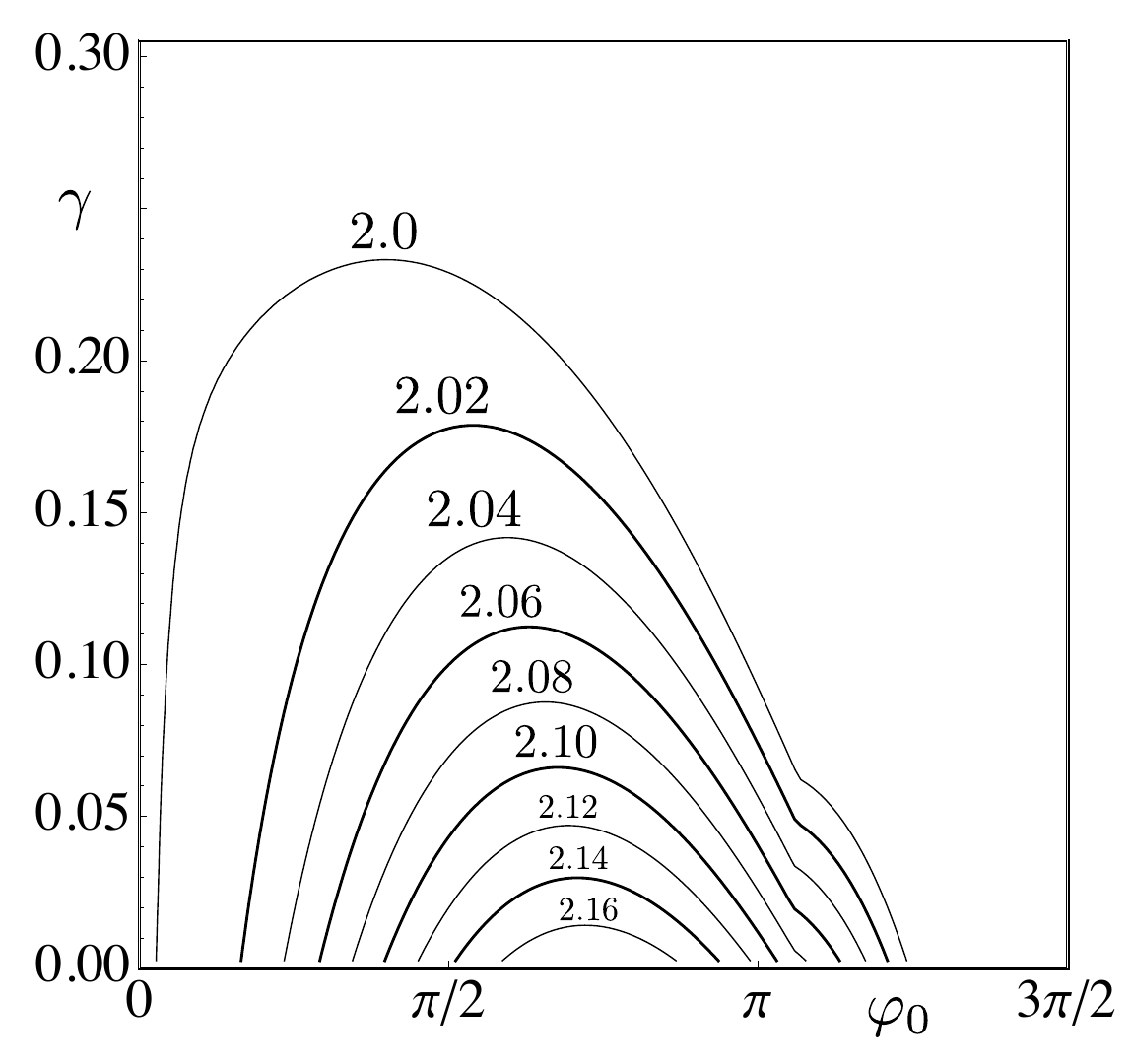}
   \caption[]
   {Bell parameter~(\ref{eq:bellfermi}) as a function of interaction parameter
   $\varphi_0$ (horizontal axis) and the width parameter $\gamma = \xi/a$
   (vertical axis) at $\eta = 0.58$.}
   \label{fig:diag4}
\end{figure}

Above, we have concentrated on the case of infinitely narrow (or high energy)
wave-packets, $\gamma = \xi/a \rightarrow 0$. The maximal value of the Bell
parameter for the general situation with a finite-width wave-packet is shown
in Fig.\ \ref{fig:diag4}, where we have chosen the optimal value $\eta = 0.58$
for narrow wave packets with $\gamma = 0$. We find that lowering the energy of
the excess electrons decreases the value of the Bell parameter and using
high-energy wave-packets happens to be the best operating limit for the Bell
test.

The interaction-equivalent phase $\chi_{10}$ involves two dimensionless
parameters, $\gamma = \xi/a$ and the Coulomb strength $\varphi_0 =
E_{12}\tau_0/\hbar$. Using phase-space arguments, one can see that these
parameters determine the number of excitations that can be created via the
wave-packet's excess energy $\delta \epsilon_\xi = \hbar v_\mathrm{F}/2\xi$ or
via the Coulomb energy $\delta \epsilon_\mathrm{C} = E_{12}$. Indeed,
multiplying these energies by the density of states $\rho = 1/\hbar
v_\mathrm{F}$ and the length $a$, we obtain the number of electrons that can
be excited within the interaction region, $N_\xi \sim a\rho \delta\epsilon_\xi
= a/2\xi$ and $N_\mathrm{C} = \delta\epsilon_\mathrm{C} \tau_0/\hbar =
\varphi_0$. The narrower the wave-packet is, the more electron-hole pairs can be
excited. On the other hand, the finite Coulomb interaction strength restricts
the number of excited particles to below $N_\mathrm{c}$, no matter how narrow
the incoming wave-packet is. As a result a further increase in the energy of
the incoming particles beyond $\delta \epsilon_\mathrm{C}$ is not harmful
since it does not produce more electron-hole excitations and thus does not
lead to a further reduction of the Bell parameter.

Decreasing the number of parasitic electron-hole pairs  should be a good
strategy to enhance the value of the Bell parameter. Keeping the interaction
fixed (say, near $\varphi_0 = \pi$), one could choose spatially extended
wave-packets such that $N_\xi$ drops below $N_\mathrm{C}$. In this situation
the number of electron-hole pairs is defined by the smaller number $N_\xi$.
However, for extended wave-packets with large $\xi$, the electrons are
unlikely to meet each other in the interaction region, thereby decreasing the
interaction phase, see Eq.\ (\ref{eq:intphase}). As a result of this
trade-off, the Bell inequality can never be violated in this regime, though
the number of parasitic electron-hole excitations is small.  However, using
extended wave-packets, one could count the particles in the outgoing leads
within a finite time window $\delta t$, $\hat{N}_i(t) = \int_{t-\delta
t/2}^{t+\delta t/2} dt\, \hat{I}_i(t)$, thus projecting the electron
trajectories on the component where the electrons have passed the interaction
region simultaneously. For $\xi > a$, the number of excited electron-hole
pairs is limited by $N_\xi \ll 1$ and one can observe a sufficient violation
of the Bell inequality.  In this approach, however, the entanglement is likely
to be induced by the projection (or the Bell measurement) itself and the setup
cannot be used as a genuine source of entangled particles.

So far, our results have been obtained for the specific form of the
interaction kernel $\kappa(x) = \exp(-|x|/a)$ with smooth tails. A different
form of $\kappa(x)$ may result in a different value of the maximal Bell
parameter. To check how strongly the result depends on the shape of
$\kappa(x)$, we consider another kernel $\kappa(x) = 1$ for $x\in[-a,a]$ and
$0$ otherwise. This kernel has sharp edges and generates an interaction
equivalent phase of the form
\begin{equation}
      \chi_{10}(t) = 4n_{10}\int d\nu\, \frac{\sin^2\nu}{\nu} 
      \frac{e^{-|\nu|\gamma +i\nu
      t/\tau_0}}{\nu - i n_{10}(1 - e^{-2i\nu})}
\end{equation}
and similarly for the phase $\chi_{01}$ with $n_{10} \rightarrow n_{01}$, see
Eq.\ (\ref{eq:phsym}). The numerical analysis for high-energy wave packets
($\gamma \to 0$) produces a maximal value of the Bell parameter ${\cal B}
\approx 2.15$ which is assumed at lower interaction $\varphi_0 \approx 0.34\,\pi$
and a slmilar value $\eta \approx 0.53$.
\begin{figure}
   \includegraphics[width=7.0cm]{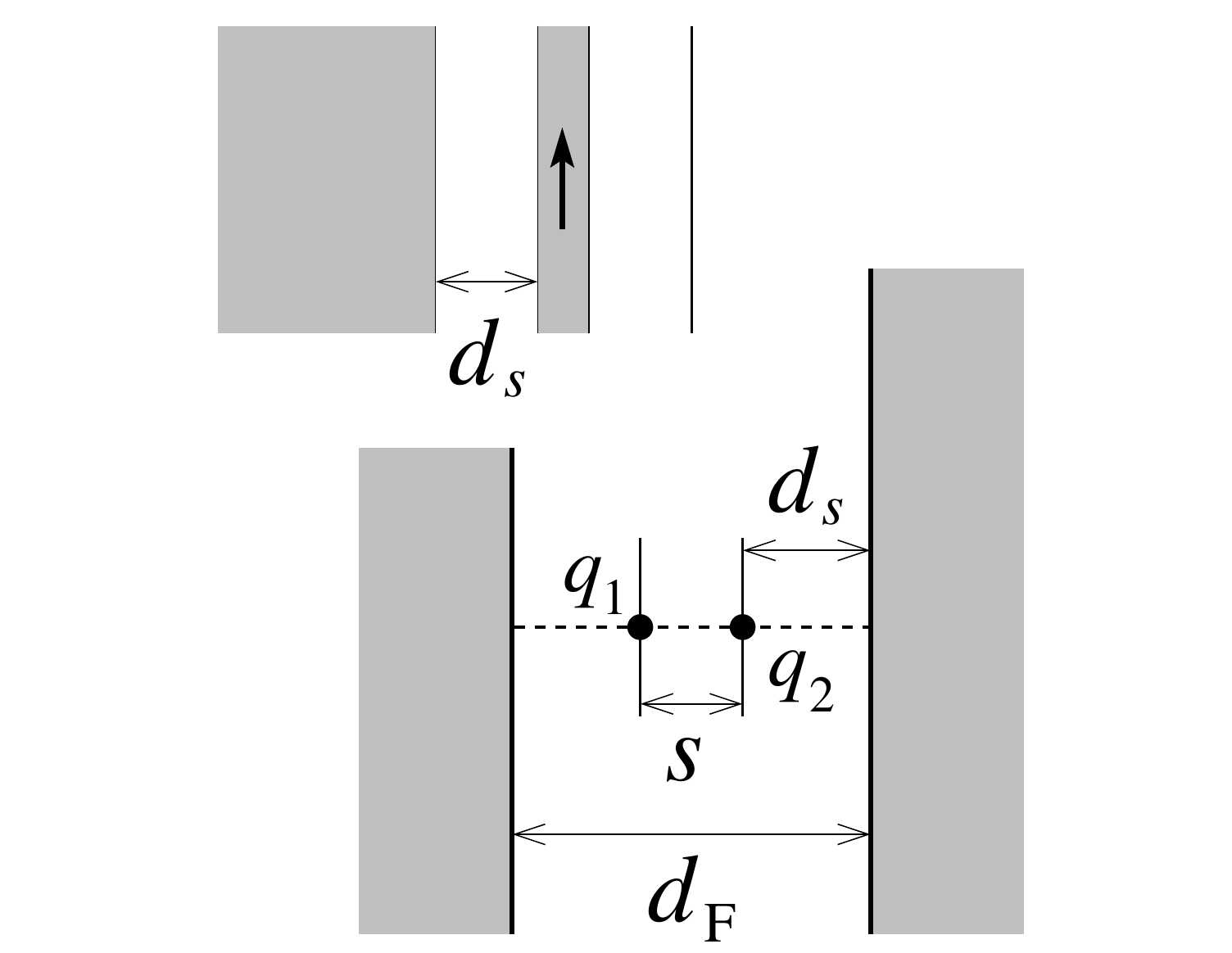}
   \caption[]
   {Interaction region between edge states `1' and `2' in a quantum Hall
   device. The edge states are separated from the closest screening Fermi
   sea by a distance $d_s \approx \ell^2/\xi$ (top left). We model the
   capacitive interaction Hamiltonian (\ref{eq:hint}) by analyzing two
   charges $q_1$ and $q_2$ separated by $s$ and screened by metallic
   plates at a distance $d_\mathrm{F}$.}
   \label{fig:edgeCap}
\end{figure}

Finally, we discuss the capacitive Hamiltonian~(\ref{eq:hint}) used throughout
the discussion. In a realistic situation the injected electrons experience a
Coulomb interaction which is screened by the Fermi sea when propagating
through the adjacent interaction leads. The peculiarity of the edge states of
the Quantum Hall system is that the edge state $\varphi_k(y)\, e^{ikx}$ with a
wave-vector $k>0$ has a transverse component
\begin{equation}
      \varphi_k(y) = H_n\Bigl( \frac{y +
      k\ell^2}\ell\Bigr)\exp\Bigl( -
      \frac{(y+k\ell^2)^2}{\ell^2}\Bigr)
\end{equation}
which is spatially separated by $k \ell^2$ from the Fermi sea electrons
occupying states with $k < 0$, see Fig.\ \ref{fig:edgeCap} (here $\ell =
\sqrt{\varphi_0/2\pi B}$ is the cyclotron length and $H_n(y)$ are Hermite
polynomials; we assume a flat boundary potential). Denoting by $d_\mathrm{F}$
the distance between the two Fermi seas, Lorentzian wave-packets with a small
width $\xi$ then can be brought close to each other, with a distance
$d_\mathrm{F}-2d_s$ between the two wave packets and a distance $d_s \approx
\ell^2/\xi$ from the nearest Fermi surface screening the interaction.

We model this situation by two parallel grounded metallic gates (accounting
for the screening effect of the Fermi seas in the adjacent interacting arms)
separated by a distance $d_\mathrm{F}$ with two charges $q_1$ and $q_2$ at a
separation $s < d_\mathrm{F}$ and located symmetrically along a line
perpendicular to the metallic plates. The electrostatic energy of this system
is given by,
\begin{equation}
      E(q_1,q_2) = \frac{E_{11}}2q_1^2 +
      E_{12}\,q_1 q_2 + \frac{E_{22}}2q_2^2,
\end{equation}
with coupling constants ($E_{11} = E_{22}$)
\begin{eqnarray} \label{eq:Eij}
      E_{ii} &=& \frac1{d_\mathrm{F}}\Bigl[ \psi\Bigl(\frac{1+\alpha}2
      \Bigr) + \psi\Bigl( \frac{1-\alpha}2\Bigr) - 2\psi(1)\Bigr],
      \\ \nonumber
      E_{12} &=& \frac1{d_\mathrm{F}}\Bigl[2\psi\Bigl(\frac12\Bigr) -\psi\Bigl(
      1 + \frac\alpha2 \Bigr) - \psi\Bigl( 1 - \frac\alpha2\Bigr)  +
      \frac1\alpha\Bigr],
\end{eqnarray}
where $\psi(x)$ is the digamma function and $\alpha = s/d_\mathrm{F}$. 
The last term $\propto 1/\alpha$ in the off-diagonal coupling strength
$E_{12}$ is the direct Coulomb interaction $q_1 q_2/s$ of the two
charges.  It is this term which allows to realize a situation where
$|E_{12}| > |E_{11}|$ when $\alpha < 0.177$.

\section{Conclusion}\label{sec:Con}

We have studied the on-demand generation of entangled electron pairs in a
double Mach-Zehnder interferometer implemented in a quantum Hall setting; the
entanglement is generated via the capacitive interaction between two
neighboring interferometer arms, shifting the phase of one wave-function
component with respect to the other.  The resulting entangled state is
analyzed in a Bell test measuring the particle numbers in the outgoing leads;
the second beam splitters and the fluxes threading the interferometers serve
to define the four measuring conditions (polarizations) in the Bell test.

Our first study with only two electrons present in the device serves to
identify the optimal conditions to generate full entanglement and maximal
violation of the Bell test in the simplest situation. We find that
simultaneous injection of high-energy pairs provides the best conditions for
entanglement; broad (low-energy) wave-packets reduce the probability that the
electrons meet and interact, as does a delay between the particles.
Furthermore, extended wave-functions are deformed in the interaction region
and thus transfer `which-path' information to the outgoing leads. Since our
Bell test is sensitive only to the particle number in the outgoing leads, this
(discarded) wave-function deformation entails a reduction of violation in the
Bell test.

Including the Fermi sea, decoherence and additional `which-path' information due
to the generation of electron-hole pairs within the interaction region are
expected to further reduce the entanglement in the relevant degrees of
freedom, the particle numbers measured in the outgoing leads.  The most
important question addressed in the present work then is: can such an
idealized device generate sufficient entanglement to be observed in a Bell
test ? We find that this is indeed the case, provided the capacitive coupling
between the two arms can be implemented such that the mutual interaction
dominantes over the self-interaction.

In our work we consider electrons in chiral states of a quantum Hall device
and a purely capacitive (in particular, non-resistive) coupling. Technically,
this allows for the bosonization of the scattering problem describing the
interaction region (however, not the scattering at the beam splitters), which
is at the heart of transforming an incoming Slater determinant into an outgoing
Slater determinant; this allows us to describe the effect of the interaction
via a voltage-pulse, thus replacing an apparent many-body problem by a
single-particle evolution.  Physically, this implies that the decoherence is
reduced in our system and limited to the creation of particle-hole pairs. It
turns out, that these particle-hole pairs further reduce (as compared with the
two-particles case) the entanglement and the violation in the Bell test, but
to a limited degree, still providing a violation ${\cal B} \approx 2.18 > 2$
in an optimized situation. Regarding the experimental implementation, we refer
the reader to the layout described in Ref.\ \onlinecite{portier}.  A favorable
finding is that the Bell test can be performed by only changing the fluxes
through the loops (possibly by using a side gate deforming the loop area) and
does not require tuning of the beam splitters. An interesting further element
is, how correction pulses resurrecting the wave function behind the scatterer
as described in Ref.\ \onlinecite{lebedev11} can help to restore a higher
degree of Bell-inequality violation.

We acknowledge financial support from the Swiss National Science Foundation
through the National Center of Competence in Research on Quantum Science and
Technology (QSIT), the Pauli Center for Theoretical Studies at ETH Zurich, and
the RFBR Grant No.\ 11-02-00744-a.

\appendix

\section{Wave packet scattering}

We show that the scattering state of two Lorentzian wave-packets $f_{1,2}(x)$
propagating along two interacting adjacent leads $`1'$ and $`2'$ is indeed
described by a Slater determinant state and thus can be prepared by applying
an evolution operator generated by a single-particle Hamiltonian.  We assume
that this single-particle evolution can be generated by voltage pulses, i.e.,
the Hamiltonian accounts for the interaction of the electrons with voltage
pulses applied at $x=0$ and adds the time dependent phases $\chi(t)$ and
$\chi'(t)$ to the state in the leads $`1'$ and $`2'$, see Eq.\
(\ref{eq:Udef}). In order to demonstrate that there are specific phases
$\chi(t)$ and $\chi'(t)$ generating the scattering state, we need to calculate
the overlap,
\begin{widetext}
\begin{equation}
      {\cal O}\! =\! \int\! dx_1 dx_2\, f_1^*(x_1) f_2^*(x_2) \!\int\! dx_1^\prime
      dx_2^\prime f_1(x_1^\prime) f_2(x_2^\prime)
      \bigl\langle\hat\Psi_2(x_2,t_0) \hat\Psi_1(x_1,t_0)
      \hat{ U}_1^\dagger[\chi] \hat{U}_2^\dagger[\chi']
      \hat{S}\hat\Psi_1^\dagger(x_1^\prime,t_0)
      \hat\Psi_2^\dagger(x_2^\prime,t_0)\bigr\rangle,
      \label{ap:O}
\end{equation}
and show, that there exist phases $\chi(t)$ and $\chi'(t)$ such that the
modulus $|{\cal O}|$ reaches unity. In Eq.\ (\ref{ap:O}), $t_0 \rightarrow
-\infty$ and the averaging is taking over zero temperature Fermi sea. In order
to calculate the overlap, we can make use of the Green's function,
\begin{eqnarray}
      {\cal C}(12|1^\prime2^\prime) = \bigl\langle T_K\{
      \hat{U}_1[\chi(t^-)]
      \hat{U}_2[\chi'(t^-)]
      \hat{S}_+\, \hat\Psi_2(x_2,t_0^-)
      \hat\Psi_1(x_1,t_0^-) \hat\Psi_1^\dagger(x_1^\prime,t_0^+)
      \hat\Psi_2^\dagger(x_2^\prime,t_0^+)\}\bigr\rangle,
      \label{ap:green}
\end{eqnarray}
\end{widetext}
where we have introduced the Keldysh time ordering $T_K$ and have defined the
fields $\chi(t)$ and $\chi'(t)$ on the lower branch of the Keldysh contour
(going back in time), while the evolution operator
\begin{equation}
      \hat{S}_+ = \hat{T}_K \exp\Bigl[ -
      \frac{i}{\hbar}\int\limits_{-\infty}^{+\infty} dt\,
      \hat{H}_\mathrm{int}(t^+) \Bigr]
\end{equation}
is defined on the upper branch of the Keldysh contour (forward in time).

We decouple the quadratic interaction Hamiltonian~(\ref{eq:hint}) with the
Hubbard-Stratonovich transform Eq.~(\ref{eq:HST}) and obtain the evolution
operator $\hat{S}_+$ in the form
\begin{eqnarray}
   \hat{S}_+ &=& \int D z_1(t)\, D z_2(t)\,  
   \hat{S}_1[\tilde{z}_1] \hat{S}_2[\tilde{z}_2] \\ \nonumber
   &&\qquad
   \times 
   \exp\Bigl[\frac{i}2 
   \int dt [\epsilon_1 z_1^2(t) + \epsilon_2 z_2^2(t)]\Bigr],
\end{eqnarray}
where the real fields $z_{1,2}(t)$ are non-zero on the upper branch of the
Keldysh contour. The actions (see Eq.\ (\ref{eq:quadr_form}) for the definition of $n_i$)
\begin{equation}
      \hat{S}_i[\tilde{z}_i] = T_K\exp\Bigl[ - i\int dt\,
      \tilde{z}_i(t) \hat{n}_i(t) \Bigr], \quad i = 1,2,
\end{equation}
with $\tilde{z}_i(t) = \epsilon_1 z_1(t)e_{1i} + \epsilon_2 z_2(t) e_{2i}$,
describe the forward in time evolution of the electrons in the leads $i = 1,2$
under the action of the time dependent fields $\tilde{z}_i(t)$ acting within
the interaction region according to the Hamiltonians $\hat{H}_{\tilde{z}_i}(t)
= \int dx\, \tilde{z}_i(t) \kappa(x) \hat\rho_i(x)$. The Green's function
(\ref{ap:green}) can be factorized with respect to the leads,
\begin{equation}
      {\cal C}(12|1^\prime2^\prime) = \bigl\langle {\cal
      C}(1|1^\prime) {\cal C}(2|2^\prime) 
      \bigr\rangle_{\tilde{z}_1,\tilde{z}_2},
      \label{ap:g2}
\end{equation}
where the average is taken over the fields $\tilde{z}_{1,2}$ and
\begin{eqnarray}\label{ap:g1}
   {\cal C}(x t_0|x^\prime t_0) &=& \bigl\langle T_K\{
   \hat{U}[\chi^-] \hat{S}[\tilde{z}^+] \\ \nonumber
   && \qquad\qquad\times
   \hat\Psi(x,t_0^-)
   \hat\Psi^\dagger(x^\prime,t_0^+)\} \bigr\rangle
\end{eqnarray}
with $\hat{S}[\tilde{z}^+]$ the action in the relevant lead.  Next, we make
use of the bosonization technique, see Ref.\ \onlinecite{delft} for details,
and express the density operator $\hat\rho(x,t)$ through the chiral bosonic
field $\hat\rho(x,t) = \partial_x \hat\theta(x-v_\mathrm{F}t)/2\pi$ where
$\hat\theta(x) = -\sum_{k>0} \bigl( \hat{b}_k e^{ikx} + \hat{b}_k^\dagger
e^{-ikx}\bigr)/\sqrt{k}$ and $\hat{b}_k^\dagger$, $\hat{b}_k$ are bosonic
creation and annihilation operators obeying the standard commutation relation
$[\hat{b}_k, \hat{b}_{k^\prime}^\dagger] = \delta_{kk^\prime}$. The electronic
field operator $\hat\Psi(x)$ can be expressed through the field
$\hat\theta(x)$ via $\hat\Psi(x) = \hat{F} e^{-i\theta(x)}/\sqrt{2\pi \delta}$
with $\delta \rightarrow 0^+$ an ultraviolet cutoff and $\hat{F}$ is a Klein
factor acting as an electron-number ladder operator.

Calculating the bosonic averages in (\ref{ap:g1}), one arrives at the result
\begin{widetext}
\begin{eqnarray}
   {\cal C}(\tau|\tau^\prime) = \frac1{2\pi i v_\mathrm{F}}
   \frac{e^{i[\chi]_+(\tau^\prime) 
   + i[\chi]_-(\tau)}}{\tau-\tau^\prime -i\delta}
   \exp\biggl\{\int\limits_0^\infty \frac{d\omega}{2\pi}
   \Bigl[ i\tilde{z}(\omega) \kappa^*(\omega)
   e^{i\omega\tau^\prime} &+& i\tilde{z}^*(\omega) \kappa(\omega) 
   \bigl( e^{-i\omega\tau} + {i\omega\chi(\omega)}/{2\pi}\bigr)
   \\ \nonumber 
   &-&|\tilde{z}(\omega)|^2\Pi_{++}(\omega) 
   - v_\mathrm{F}^2 |\chi(\omega)|^2G_{++}^*(\omega) \Bigl] \biggr\},
\end{eqnarray}
where we have introduced retarded variables $\tau = t_0 -x/v_\mathrm{F}$ and
$\tau^\prime = t_0 - x^\prime/v_\mathrm{F}$; the quantities $G_{++}(\omega)$
and $\Pi_{++}(\omega)$ are defined in Eq.\ (\ref{eq:rr}). Substituting this
expression into Eq.~(\ref{ap:g2}) and taking the Gaussian integrals over the
fields $z_1(t)$ and $z_2(t)$, one finally arrives at the following expression
for the correlation function ${\cal C}(12|1^\prime2^\prime)$,
\begin{eqnarray}
   &&{\cal C}(12|1^\prime2^\prime) =
   \frac{e^{i[\chi]_+(\tau_1^\prime) + i[\chi]_-(\tau_1) +
   i[\chi']_+(\tau_2^\prime) + i[\chi']_-(\tau_2)}}{
   (2\pi i v_\mathrm{F})^2(\tau_1-\tau_1^\prime - i\delta)
   (\tau_2-\tau_2^\prime-i\delta)}
   \exp\Bigl[ -v_\mathrm{F}^2\int_0^\infty \frac{d\omega}{2\pi}
   \,\bigl(|\chi(\omega)|^2 + |\chi'(\omega)|^2\bigr) G_{++}^*(\omega)
   \Bigr]
      \\ &&\nonumber
   \times \prod_{k=1,2}
   \exp\Bigl[ -i\int_0^\infty \frac{d\omega}{2\pi}\,
   F_k^*(\omega)\bigl( e_{k1}e^{i\omega\tau_1^\prime} +
   e_{k2}e^{i\omega\tau_2^\prime}\bigr)\Bigl(e_{k1}\bigl(
   e^{-i\omega\tau_1} + {i\omega\chi(\omega)}/{2\pi}\bigr)
   + e_{k2}\bigl( e^{-i\omega\tau_2} +
   {i\omega\chi'(\omega)}/{2\pi}\bigr)\Bigr)\Bigr],
\end{eqnarray}
\end{widetext}
with the functions $F_k(\omega)$, $k=1,2$, defined in Eq.\ (\ref{eq:F}).  This
correlator is an analytic function of the retarded variables $\tau_1$ and
$\tau_2$ ($\tau_1^\prime$ and $\tau_2^\prime$) in the lower (upper) half
plane.  This feature allows us to evaluate the integrals over the coordinates
in Eq.\ (\ref{ap:O}) involving Lorentzian wave-packets with the widths $\xi_1$
and $\xi_2$. In the end, we arrive at the following expression for the overlap
\begin{widetext}
\begin{eqnarray}
      &&{\cal O} = \exp\Bigl[
      i[\chi]_+(i\tau_{\xi_1}) +
      i[\chi]_-(-i\tau_{\xi_1})+
      i[\chi']_+(i\tau_{\xi_2})+
      i[\chi']_-(-i\tau_{\xi_2})
      -v_\mathrm{F}^2\int_0^\infty \frac{d\omega}{2\pi}
      \bigl( |\chi(\omega)|^2 +|\chi'(\omega)|^2 \bigr)
      G_{++}^*(\omega)\Bigr]
      \nonumber\\
      &&\times\prod_{k=1,2}
      \exp\Bigl[ -i\int_0^\infty \frac{d\omega}{2\pi}\,
      F_k^*(\omega)\bigl( e_{k1}e^{-\omega\tau_{\xi_1}} +
      e_{k2}e^{-\omega\tau_{\xi_2}}\bigr)\Bigl(e_{k1}\bigl(
      e^{-\omega\tau_{\xi_1}} + {i\omega\chi(\omega)}/{2\pi}\bigr)
      + e_{k2}\bigl( e^{-\omega\tau_{\xi_2}} +
      {i\omega\chi'(\omega)}/{2\pi}\bigr)\Bigr)\Bigr].
      \nonumber
\end{eqnarray}
\end{widetext}
Minimizing the real part of the overall exponential provides the optimal
fields $\chi$ and $\chi'$ in Eqs.\ (\ref{eq:phiup}) and (\ref{eq:phidn})
producing a maximal overlap $|{\cal O}| = 1$; the overall phase is given by
Eq.\ (\ref{eq:phase}).


\begin{thebibliography}{99}
\bibitem{driftvel} In the Hall-bar geometry, the chiral electrons move (along
$x$) in the transverse potential ($U(y)$) generated by the gates and the Fermi
velocity $v_\mathrm{F}$ should be replaced by the (local) drift velocity
$v_\mathrm{drift} \sim -\ell^2 [\partial_y U]_{y_0}/\hbar$, with $y_0$ the
position of the edge channel.
\end{thebibliography}
\end{document}